\documentclass{jfm}
\usepackage{graphicx}
\usepackage{epstopdf, epsfig}

\usepackage{hyperref}

\usepackage{bm}
\usepackage{float}
\usepackage{caption}
\usepackage{subcaption}
\usepackage[title]{appendix}
\usepackage{amsmath,amssymb,amsfonts}
\usepackage{graphicx}
\usepackage{epstopdf}
\usepackage[dvipsnames]{xcolor}
\usepackage{comment}
\usepackage[overload]{empheq}
\usepackage{upgreek}
\usepackage{lmodern,pdftexcmds}
\usepackage[numbers,sort&compress]{natbib}
\usepackage{cases}
\usepackage{nccmath}
\usepackage{scalerel}
\usepackage{color,soul}
\usepackage{wasysym}
\usepackage[makeroom]{cancel}
\usepackage{hyperref}
\usepackage[mathlines]{lineno}
\usepackage{dcolumn}
\usepackage{changepage}
\usepackage{lipsum}

\renewcommand{\bm}[1]{\boldsymbol{\mathbf{#1}}}

\providecommand*{\bm}{\mathbf}

\newcommand{\bv}[1]{\ensuremath{\boldsymbol{#1}}} 
\newcommand{\gv}[1]{\ensuremath{\mbox{\boldmath$ #1 $}}} 

\renewcommand{\O}{\mathcal{O}}
\newcommand\bsigma{\bm{\sigma}}  

\shorttitle{Density-contrast induced inertial forces on particles in oscillatory flows}
\shortauthor{S. Agarwal, G. Upadhyay, Y. Bhosale, M. Gazzola and S. Hilgenfeldt}

\title{Density-contrast induced inertial forces on particles in oscillatory flows}

\author{Siddhansh Agarwal\aff{1}\footnote{Present Address: Department of Bioengineering, University of
California, Berkeley, CA 94720, USA}, Gaurav Upadhyay\aff{1},
  Yashraj Bhosale\aff{1}, Mattia Gazzola\aff{1,4}
 \and Sascha Hilgenfeldt\aff{1}\corresp{\email{sascha@illinois.edu}}}

\affiliation{\aff{1}Department of Mechanical Science and Engineering, University of Illinois,
Urbana Champaign, IL 61801, USA
\aff{2}Carl R.\ Woese Institute for Genomic Biology, University of Illinois, Urbana-Champaign, IL 61801, USA}

\allowdisplaybreaks

\begin{document}

\maketitle
\allowdisplaybreaks
\begin{abstract}
Oscillatory flows have become an indispensable tool in microfluidics, inducing  inertial effects for displacing and manipulating fluid-borne objects in a reliable, controllable, and label-free fashion. However, the quantitative description of such effects has been confined to limit cases and specialized scenarios. 
Here we develop an analytical formalism yielding the equation of motion of density-mismatched spherical particles in arbitrary background flows, generalizing previous work. Inertial force terms are systematically derived from the geometry of the flow field together with analytically known Stokes number dependences.  
Supported by independent, first-principles direct numerical simulations, we find that these forces 
are important even for nearly density-matched objects such as cells or bacteria, enabling their fast displacement and separation. Our formalism thus generalizes the Maxey--Riley equation, encompassing not only particle inertia, but consistently recovering, in the limit of large Stokes numbers, the Auton 
modification to added mass as well as
the far-field acoustofluidic secondary radiation force. 
\end{abstract}


\section{Introduction}

One of the most fundamental problems in fluid dynamics that has evaded a general solution is describing the motion of particles immersed in a prescribed background flow. Most analytical attempts work under the severe assumption of reversible unsteady Stokes flows, for which symmetry-breaking inertial effects  are neglected
(see \cite{michaelides1997transient} for a brief overview). It was the seminal work by \citet{maxey1983equation} (MR) that first characterized, rigorously and systematically,  hydrodynamic forces on particles, albeit strictly in the limit of vanishing inertial effects. 
As a result, 
the MR equation has been used extensively over the last forty years. 

The MR equation (for spherical particles) assumes the validity of the  unsteady Stokes assumption, which implies that
(i) the particle Reynolds number based on a typical {\em difference} velocity between particle speed and background flow must be small, and (ii) the background flow gradients must be small compared to viscous momentum diffusion. 
These assumptions do constrain the applicability of MR in a number of situations. 
One of the most glaring shortcomings was pointed out by \cite{leal1992laminar}, and concerns the incompatibility of MR with the experimentally observed phenomenon of lateral migration of particles due to lift forces caused by inertial effects. 
Subsequent work aimed at the development of equations valid at finite particle Reynolds numbers has yielded specialized results, for example for steady flow \citep{ho1974inertial,martel2014inertial,hood2015inertial} or for forces occurring in acoustic fields \citep{baudoin2020acoustic,rufo2022acoustofluidics}.


The advent of oscillatory microfluidics \citep{marmottant2003controlled,thameem2017fast,lut03,zhang2020acoustic,zhang2021versatile,mutlu2018oscillatory,zhang2021portable,zhang2023sonorotor} has since introduced the use of much stronger particle inertia effects, enabling fast and high-throughput particle manipulation. Yet again, quantitative modeling and prediction of such effects has been largely lacking, with experimental results often explained qualitatively, and/or by appealing to specialized theories such as acoustofluidics \citep{chen2016onset,devendran2014separation,collins2019acoustic,wu2019acoustofluidic}. Given the versatility and richness of microfluidic flows, what is needed is a fundamental understanding of inertial hydrodynamic forces acting on particles immersed in a general unsteady background flow, that is, a true generalization of MR.


In a first step towards such a generalization, \citet{agarwal2021unrecognized} rigorously described 
inertial forces on density-matched particles.
Whereas MR does not predict any net force on neutrally buoyant particles immersed in unsteady fluid flows,  \citet{agarwal2021unrecognized}  showed that such a force can be very significant and is often dominant in oscillatory microfluidics. In the present paper, we augment that formalism to include finite density contrast between particle and fluid (a relevant scenario in microfluidics), thus completing the consistent generalization of MR.
In our theory, density-contrast dependent contributions to inertial forces specialize to the well-known \cite{auton1988force} correction in the potential flow limit, but continue to play a significant role in the presence of unsteady viscous effects.
In a different limit our framework recovers acoustofluidic formulae for radiation forces on particles, while again incorporating viscous effects quantitatively.

The organization of this paper is as follows. In Sec.~\ref{sec:theory} we describe the general theoretical formalism for inertial forces and their  evaluation for oscillatory flows.
In Sec,~\ref{sec:eom}, we develop an explicit time-averaged equation of motion for spherical particles, and in Sec.~\ref{sec:results} we rigorously compare its predictions with direct numerical simulations as well as with existing theories in specialized limits. Section \ref{sec:disc} discusses the validity and importance of the present approach in practical situations, while Sec.~\ref{sec:concl} draws conclusions.

\section{Theoretical Formalism}\label{sec:theory}
\subsection{Problem set-up}
\begin{figure}
		\centering
		\includegraphics[width=\textwidth]{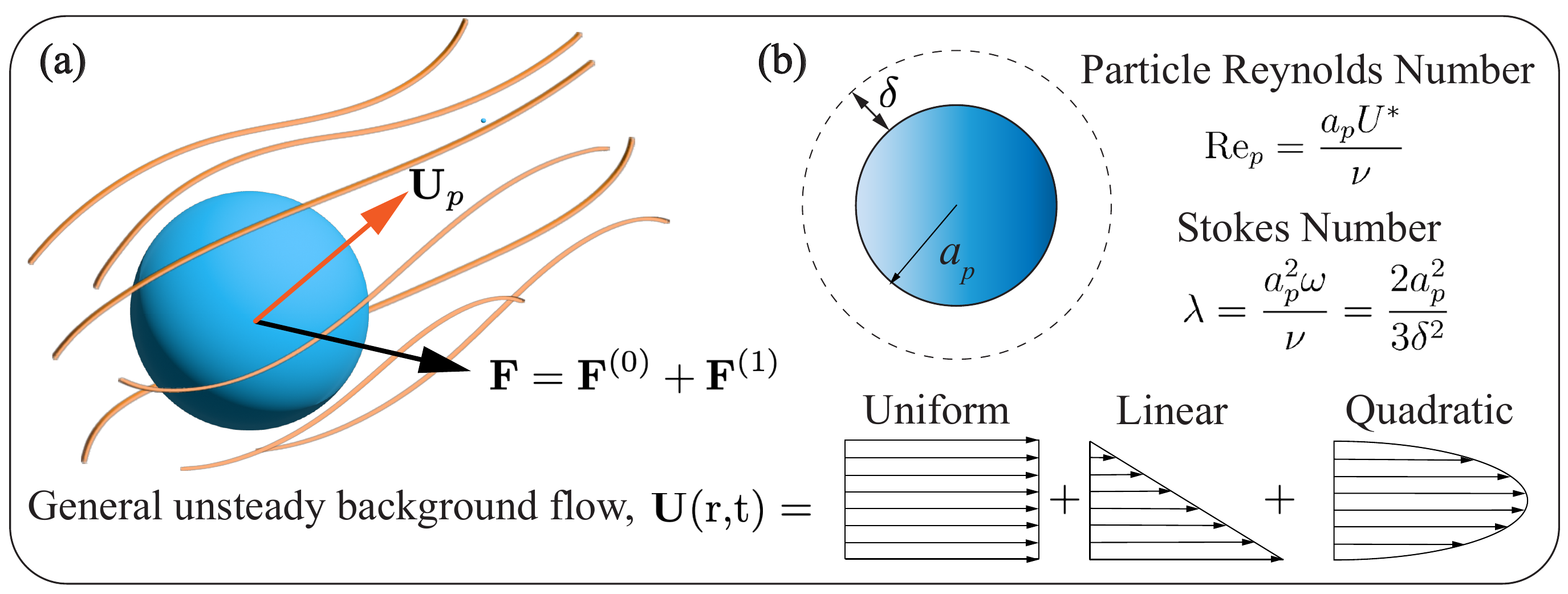}
		\caption{{\bf (a)} Schematic of a spherical particle of radius $a_p$ moving with a velocity $\bm{U}_p$ as a consequence of the hydrodynamic force $\bm{F}$ exerted by the surrounding fluid. The undisturbed flow field far away from the particle is denoted by $\bm{U}$. The hydrodynamic force is generally decomposed into a force due to the undisturbed flow $\bm{F}^{(0)}$ and the disturbance flow $\bm{F}^{(1)}$ due to the presence of the particle. {\bf (b)} The unsteadiness of the flow introduces the Stokes number $\lambda$, which, for oscillatory flows, is a function of the ratio of the particle size to the oscillatory boundary layer thickness $\delta$. The background flow is Taylor expanded around the particle center up to the quadratic term.} \label{fig:theory}
\end{figure}
In this paper we develop a unifying theory for the equation of motion of spherical particles of radius $a_p$ and mass $m_p$ immersed in general unsteady incompressible Newtonian flows of fluid density $\rho_f$ (Fig.~\ref{fig:theory}), placing particular emphasis on fast oscillatory flows, while consistently accounting for particle inertia. 
The characteristic speed $U^*$ of the unsteady background flow and kinematic viscosity $\nu$ of the fluid define the particle Reynolds number $\operatorname{Re}_{p}=a_p U^*/\nu$.
The particle reaction to the flow importantly also depends on the Stokes number, which we define as 
 $\lambda=a_p^2\omega/(3\nu)$.
 The unsteady time scale of the flow is written as $1/\omega$, anticipating the oscillatory case, where we can alternatively use the Stokes boundary layer scale $\delta=(2\nu/\omega)^{1/2}$ to write $\lambda=2a_p^2/3\delta^2$.
 In oscillatory microfluidics, 
 we typically have $\lambda\sim 1-10$, while acoustofluidics generally operates at $\lambda \gg 1$. The case $\lambda\ll 1$ is usually not practically relevant, as the resulting inertial forces on particles become very small. 

We follow MR in decomposing the flow around the particle into the given background flow  $\bm{U}$ present without the particle, and the disturbance flow due to the particle's presence. Forces caused by the background flow will carry the superscript $^{(0)}$, while those stemming from the disturbance flow will have the superscript $^{(1)}$. All forces will be computed for arbitrary $\lambda$ and to first order in $Re_p$, using a regular perturbation expansion. In general flows, such an approach is valid in an inner region, while an outer region (in which inertia reasserts dominance) would have to be treated separately and the complete problem solved by asymptotic matching. However, as shown by \cite{lovalenti1993hydrodynamic}, an outer region is not present when 
the oscillatory inertia of the disturbance flow is much greater than its advective inertia. We quantify below (see Sec.~\ref{sec:outerflow}) that this criterion is comfortably fulfilled for practically relevant flows in oscillatory microfluidics, so that it is sufficient to demonstrate the solution by regular expansion.

\subsection{Particle motion and fluid flow}
Our task is thus to determine explicit expressions of terms in the following equation of motion for the particle velocity $\bm{U}_p$,
	\begin{align}
		m_p \frac{d \bm{U}_p}{dt} &= \bm{F}^{(0)}+\bm{F}^{(1)} =\bm{F}^{(0)}_0+\operatorname{Re}_p\bm{F}^{(0)}_1+\bm{F}^{(1)}_0+\operatorname{Re}_p\bm{F}^{(1)}_1 + \dots\,,\label{eommaster}
	\end{align}
where subscripts denote orders of $\operatorname{Re}_{p}$. Note that the decomposition of $\bm{F}^{(0)}$ is exact (there are no terms of higher order in $\operatorname{Re}_{p}$, cf.~\cite{maxey1983equation}), while we truncate the expansion of $\bm{F}^{(1)}$ at first order.
Expressions for $\bm{F}^{(0)}_0$, $\bm{F}^{(1)}_0$, and part of $\bm{F}^{(0)}_1$ are contained in MR, and we determine the remaining terms here. 

Hydrodynamic force components are computed from the flow field stresses, as $\bm{F}^{(i)}=(F_S/6\pi)\oint_S \bm{n}\cdot \bsigma^{(i)} dS$ with $i=0,1$, where we use the Stokes drag scale $F_S/6\pi=\nu \rho_f a_p U^*$, and the integral is over the particle surface with its outward normal $\bm{n}$. We use lowercase letters for velocities non-dimensionalized by $U^*$ and it is advantageous in intermediate results to evaluate these velocities in a coordinate system moving with the particle center, writing 
$\boldsymbol{w}^{(0)}=\bm{u}-\bm{u}_p$ for the undisturbed background flow and $\boldsymbol{w}^{(1)}$ for the disturbance flow. The dimensionless fluid stress tensors are thus written $ \bsigma^{(i)}=-p^{(i)}\bm{I} + \nabla \boldsymbol{w}^{(i)} + \left(\nabla \boldsymbol{w}^{(i)}\right)^T$.

The Navier-Stokes equations in the particle frame of reference can be decomposed into background and disturbance components exactly (without approximations),
\begin{subequations}
	\begin{align}
			\nabla^{2} \boldsymbol{w}^{(0)}-\nabla p^{(0)} =& 3\lambda \frac{\partial \boldsymbol{w}^{(0)}}{\partial t}+\operatorname{Re}_{p}\left(\boldsymbol{w}^{(0)} \cdot \nabla\boldsymbol{w}^{(0)}\right), \quad \boldsymbol{\nabla} \cdot \boldsymbol{w}^{(0)} =0,\label{undistflow} \\
   \boldsymbol{w}^{(0)} =& \bm{u}-\bm{u}_p \quad \text { as } r \rightarrow \infty, \label{undistbc}\\
			\nabla^{2} \boldsymbol{w}^{(1)}-\nabla p^{(1)} =& 3\lambda \frac{\partial \boldsymbol{w}^{(1)}}{\partial t}\nonumber\\  
   &+\operatorname{Re}_{p}\bigg[\boldsymbol{w}^{(0)} \cdot \nabla \boldsymbol{w}^{(1)}+\boldsymbol{w}^{(1)}\cdot \nabla \boldsymbol{w}^{(0)} +\boldsymbol{w}^{(1)}\cdot \nabla\boldsymbol{w}^{(1)}\bigg],\\ 
   \boldsymbol{\nabla} \cdot \boldsymbol{w}^{(1)} =& 0, \label{distflow}\\
			\boldsymbol{w}^{(1)} =& \bm{u}_p-\bm{U} \quad \text { on } r=1 \quad \text{and} \quad \boldsymbol{w}^{(1)} = 0 \quad \text { as } r \rightarrow \infty. \label{distbc}
	\end{align}\label{navierstokes}\nolinebreak
\end{subequations}

\subsection{Forces from background flow}\label{sec:backgndforces}
Both the ${\cal O}(1)$ and ${\cal O}(\operatorname{Re}_{p})$ components of $\bm{F}^{(0)}$ in \eqref{eommaster} can be evaluated directly using the divergence theorem and the above Navier-Stokes equations valid for the background flow $\boldsymbol{w}^{(0)}$. In lab coordinates 
(see \cite{maxey1983equation,rallabandi2021inertial}) they read
\begin{align}
 	\bm{F}^{(0)}_0= \frac{F_S}{6\pi} \int_{V} \left( 3\lambda \partial_t \bm{u}\right) dV,\quad
 	\bm{F}^{(0)}_1=\frac{F_S}{6\pi} \int_{V} \left(\bm{u}\cdot \nabla \bm{u}\right) dV.\label{eq:F0}
\end{align}
To make further progress, we need to evaluate forces due to successive orders of $\boldsymbol{w}^{(1)}$, which are ultimately also derived from the given background field $\bm{u}$. To render our solution strategy analytically tractable, we expand  $\bm{u}$ around the leading-order particle position $\bm{r}_{p_0}$ into spatial moments of alternating symmetry,
\begin{align}
	\bm{u}=\bm{u}|_{\bm{r}_{p_0}} + \bm{r}\cdot \bm{E} + \bm{r}\bm{r}:\bm{G}+\dots,\label{eqn:uexp}
\end{align}
where $\bm{E}=(a_p/L_\Gamma)\nabla \bm{u}|_{\bm{r}_{p_0}}$ and $\bm{G}=\frac{1}{2}(a_p^2/L_\kappa^2)\nabla\nabla \bm{u}|_{\bm{r}_{p_0}}$ are time-dependent, with gradient $L_\Gamma$ and curvature $L_\kappa$ length scales. Such an expansion is valid for $a_p/L_\Gamma\ll 1$ and $a_p/L_\kappa \ll 1$, conditions readily satisfied in microfluidic scenarios. Based on this, Eq.~\eqref{eq:F0} was recently evaluated analytically by \cite{agarwal2021unrecognized} and \cite{rallabandi2021inertial}, showing that 
an $O(\operatorname{Re}_p)$ contribution from $\bm{F}^{(0)}_1$ had been missed in MR, while being, in fact, of the same order as other terms in the original MR equation. 

\subsection{Disturbance flow: zeroth order}
The Navier-Stokes equations for the disturbance flow at $O(\operatorname{Re}_p^0)$ read
\begin{subequations}
	\begin{align}
			\nabla^{2} \boldsymbol{w}_0^{(1)}-\nabla p_0^{(1)} =&3\lambda \frac{\partial \boldsymbol{w}_0^{(1)}}{\partial t}, \quad \boldsymbol{\nabla} \cdot \boldsymbol{w}_0^{(1)} =0, \label{distflow0} \\
			\boldsymbol{w}_0^{(1)} =& \bm{u}_{p_0}-\bm{u} \quad \text { on } r=1 \quad \text{and} \quad \boldsymbol{w}_0^{(1)} = 0 \quad \text { as } r \rightarrow \infty.
	\end{align}\nolinebreak \label{nastdist0}
\end{subequations}
Unlike MR, where the solution to this unsteady Stokes equation (\ref{nastdist0}) was not explicitly needed to compute the force resulting from it, 
our present approach does require expressions for 
$\boldsymbol{w}^{(1)}_0$ to compute the full $\O(\operatorname{Re}_p)$ force. This is accomplished 
by substituting the expansion \eqref{eqn:uexp} into (\ref{nastdist0}). Each spatial moment gives rise to a linear equation with known solutions, the sum of which yields the general expression 
(see \cite{landau1959course,pozrikidis1992boundary})
	\begin{align}
	\boldsymbol{w}^{(1)}_0  =  \bm{\mathcal{M}}_D \cdot \bm{u}_s  - \bm{\mathcal{M}}_Q \cdot\left(\bm{r}\cdot \bm{E}\right) -\bm{\mathcal{M}}_O \cdot\left(\bm{r}\bm{r}:\bm{G}\right) + \dots, \label{w10gen}
\end{align}
where $\bm{u}_s = \bm{u}_{p_0}-\bm{u}\vert_{\bm{r}_{p_0}}$ is the slip velocity and $\bm{\mathcal{M}}_{D,Q,O}(r,\lambda)$ are mobility tensors with known spatial dependence. For oscillatory flows, their dependence on the Stokes number $\lambda$ is known analytically. Explicit expressions for these tensors are given in  Appendix \ref{appendix:mobility}.

\subsection{Disturbance flow: first order using a reciprocal theorem}
Fast oscillatory particle motion can give rise to large disturbance flow gradients, so that terms involving $\nabla\boldsymbol{w}^{(1)}$ on the RHS of \eqref{distflow} are not necessarily negligible compared to the viscous diffusion term on the LHS, and  $\O(\operatorname{Re}_p)$ force terms in $\bm{F}^{(1)}$ become important.

With $\boldsymbol{w}^{(1)}_0$ explicitly known, the equations at $\mathcal{O}(\operatorname{Re}_p)$ read
	\begin{subequations}
		\begin{align}
			\nabla^{2} \boldsymbol{w}^{(1)}_1-\nabla p^{(1)}_1 &=\nabla \cdot \bm{\sigma}^{(1)}_1=3\lambda \frac{\partial \boldsymbol{w}^{(1)}_1}{\partial t}+\bm{f}_0, \quad
			\boldsymbol{\nabla} \cdot \boldsymbol{w}^{(1)}_1 =0, \\
			\boldsymbol{w}^{(1)}_1 &= -\bm{u}_{p_1} \quad \text { on } r=1 \quad \text {and} \quad
			\boldsymbol{w}^{(1)}_1 = 0 \quad \text { as }\quad r \rightarrow \infty \,,
	\end{align}\label{orep}\nolinebreak
\end{subequations}
with $\bm{f}_0=\boldsymbol{w}^{(0)}_0\cdot \nabla \boldsymbol{w}^{(1)}_0+\boldsymbol{w}^{(1)}_0\cdot \nabla \boldsymbol{w}^{(0)}_0 +\boldsymbol{w}^{(1)}_0\cdot \nabla\boldsymbol{w}^{(1)}_0$ as the leading-order nonlinear forcing.

In order to compute the force $\bm{F}^{(1)}_1$, we do not solve for the flow field $\boldsymbol{w}^{(1)}_1$ in \eqref{orep} but instead employ a reciprocal relation in the Laplace domain. The reciprocal theorem infers the force from the known stress of a test flow $\bm{u}'$, which is here chosen to be a dipolar unsteady Stokes flow around the particle with arbitrary directionality $\bm{e}$---see  Appendix \ref{appendix:reciprocal} for a detailed derivation. 
We obtain for the magnitude of the force along ${\bm e}$:
	\begin{align}
		\bm{e}\cdot \bm{F}^{(1)}_1 = \frac{F_S}{6\pi}\mathcal{L}^{-1}\left\{\int_{S_p}\frac{\hat{\bm{u}}_{p_1}}{\hat{u}'}\cdot ( \hat{\bm{\sigma}}'\cdot \bm{n})dS - \operatorname{Re}_p\int_V \frac{\hat{\bm{u}}' \cdot \hat{\bm{f}}_0}{\hat{u}'}dV \right\}\,,\label{eqn:F11}
	\end{align}
where the hat denotes the Laplace transform and $\mathcal{L}^{-1}$ is the inverse Laplace transform. When applied at $O(1)$, this reciprocal-theorem strategy similarly yields 
\begin{align}
		\bm{e}\cdot \bm{F}^{(1)}_0 =F^{(1)}_0=& \frac{F_S}{6\pi}\mathcal{L}^{-1}\left\{\int_{S_p}\frac{\hat{\boldsymbol{w}}_0^{(0)}}{\hat{u}'}\cdot ( \hat{\bm{\sigma}}'\cdot \bm{n})dS\right\}\label{eqn:F10},
\end{align}
which is precisely the force expression obtained by MR. Since the variable in the overall equation of motion (\ref{eommaster}) is the unexpanded particle velocity $\bm{u}_{p}$, we make the substitution $\boldsymbol{w}^{(0)}_0=\boldsymbol{w}^{(0)} - \operatorname{Re}_p \bm{u}_{p_1}+\mathcal{O}(\operatorname{Re}_p^2)$. Adding \eqref{eqn:F10} and \eqref{eqn:F11}, the $\mathcal{O}(\operatorname{Re}_p)$ term in \eqref{eqn:F10} exactly cancels the first term in \eqref{eqn:F11} and produces a correction term that is $\mathcal{O}(\operatorname{Re}_p^2)$. The net force on the particle due to its disturbance flow then reads
\begin{subequations}
\begin{align}
	 \bm{e}\cdot \bm{F}^{(1)} = & \,\bm{e}\cdot \left(\bm{F}^{(1)}_0+\operatorname{Re}_p\bm{F}^{(1)}_1\right) +\mathcal{O}(\operatorname{Re}_p^2),\\
    \bm{e}\cdot \bm{F}^{(1)}_0= & \,\frac{F_S}{6\pi}\mathcal{L}^{-1}\left\{\int_{S_p}\frac{\hat{\boldsymbol{w}}^{(0)}}{\hat{u}'}\cdot ( \hat{\bm{\sigma}}'\cdot \bm{n})dS\right\}\\
    \bm{e}\cdot \bm{F}^{(1)}_1= & \,\frac{F_S}{6\pi}\mathcal{L}^{-1}\left\{- \operatorname{Re}_p\int_V \frac{\hat{\bm{u}}' \cdot \hat{\bm{f}}_0}{\hat{u}'}dV \right\},\label{eqn:F1}
\end{align}
\end{subequations}
where we have also replaced $\boldsymbol{w}^{(0)}_0$ by $\boldsymbol{w}^{(0)}$ in $\bm{f}_0$, resulting in an error that is again $\mathcal{O}(\operatorname{Re}_p^2)$.  Only certain products in $\bm{f}_0$ are non-vanishing when the angular integration is performed due to alternating symmetry of terms in the background flow field multipole expansion \eqref{eqn:uexp}. These non-zero terms are conveniently labeled by the multipole orders involved in the product:
\begin{align}
 \operatorname{Re}_p\frac{F_S}{6\pi}\mathcal{L}^{-1}\left\{-\int_V \frac{\hat{\bm{u}}' \cdot \hat{\bm{f}}_0}{\hat{u}'}dV\right\}=& \,F_{\sigma \Gamma}^{(1)}+ F_{\Gamma\kappa}^{(1)} + \dots.\label{eqn:F11gen}
\end{align}
Here, $F_{\sigma \Gamma}^{(1)}$, $F_{\Gamma \kappa}^{(1)}$ are the inertial force contributions obtained by successive contractions of adjacent tensors involving $\bm{u}_s$ (index $\sigma$), $\bm{E}$ (index $\Gamma$), $\bm{G}$ (index $\kappa$) and so on. The volume integral is tedious but straightforward to compute since all the integrations resulting from the leading-order velocity fields \eqref{eqn:uexp},\eqref{w10gen} are convergent. 
The evaluation of the Laplace transforms can be performed analytically if the flow has harmonic time dependence. This is not a severe restriction as arbitrary time dependences can be decomposed into harmonic contributions. To simplify notation, we therefore assume a single oscillatory frequency $\omega$ in the following, without loss of generality.

When the particle is neutrally buoyant, the first term $F_{\sigma \Gamma}^{(1)}$ vanishes so that the leading term is $F_{\Gamma \kappa}^{(1)}$, which was derived in \cite{agarwal2021unrecognized} as an unexpected inertial force for density-matched particles. This term (involving the product $\bm{E : G}$) has no analog in previous literature and for completeness, we reproduce it here for harmonic oscillatory flows $\bm{U}$:
\begin{align}
    F_{\Gamma \kappa}^{(1)}=m_f a_p^2 \left[\nabla\bm{U}:\nabla\left(\nabla \bm{U}\right)\right]\cdot \bm{e} \,\mathcal{F}^{(1)}_1\,. \label{F11F11}
\end{align}
The $\lambda$-dependent dimensionless function $\mathcal{F}^{(1)}_1$ results from the volume integration, which also yields $m_f$, the mass of fluid displaced by the particle,
via $(4\pi a_p^3/3)\operatorname{Re}_p F_S/(6\pi) = m_f a_p^2 (U^*)^2 $. 
In the next section, we follow a similar strategy for non-neutrally buoyant particles. 


\subsection{Disturbance flow: Evaluation of $F_{\sigma \Gamma}^{(1)}$}\label{sec:eval}
Non-neutrally buoyant particles have a slip velocity and thus a non-zero $F_{\sigma \Gamma}^{(1)}$, involving the product $\bm{u_s \cdot E}$. Appropriate to fast harmonic oscillatory flows, we approximate the background flow as a potential flow with  a given single frequency. The slip velocity as a linear response can then be generally decomposed into an in-phase and an out-of-phase component with respect to the background flow, i.e.,  $\bm{u}_s(\bm{r}_p,t)=\bm{u}_{s}^{I}(\bm{r}_p,t)+\bm{u}_{s}^{O}(\bm{r}_p,t)$.
The corresponding force is written as
\begin{align}
 	\frac{F_{\sigma \Gamma}^{(1)}}{\operatorname{Re}_p F_S/(6\pi)}=\frac{4\pi}{3}(\bm{u}_{s}^{I}\cdot\bm{E}) \cdot \bm{e}\, \mathcal{{G}}_1(\lambda) +  \frac{4\pi}{3}(\bm{u}_{s}^{O}\cdot\bm{E}) \cdot \bm{e}\, \mathcal{{G}}_2(\lambda) \,,\label{F11G1G2gen}
 \end{align}
where the $\mathcal{{G}}_1$ and $\mathcal{{G}}_2$ terms are explicit outcomes of the volume integration in \eqref{eqn:F11gen} and capture the $\lambda$-dependence of the in- and out-of-phase contributions, respectively. 
For fast oscillatory background flows, we can replace the in-phase component with $\bm{u}_{s}\cdot\bm{E}$ and the out-of-phase component with $\partial_t \bm{u}_{s}\cdot\bm{E}$ (see Appendix \ref{appendix:G1G2} for details), resulting in 
\begin{align}
     F_{\sigma \Gamma}^{(1)} = &\frac{4\pi}{3}\rho_f a_p^2 U^{*^2} \left( \bm{u}_s\cdot\bm{E} \cdot \bm{e}\, \mathcal{G}_1(\lambda) +  \partial_t{\bm{u}_s}\cdot\bm{E}  \cdot \bm{e}\, \mathcal{G}_2(\lambda) \right)\nonumber\\
    =& m_f \left[ (\bm{U}_p -\bm{U})\cdot \nabla \bm{U}\right]\cdot \bm{e}\, \mathcal{G}_1(\lambda) + m_f \left[ \partial_t(\bm{U}_p -\bm{U}) \cdot \nabla \bm{U}\right] \cdot \bm{e}\, \frac{\mathcal{G}_2(\lambda)}{\omega}.\label{Fsiggam}
\end{align}

While the exact, lengthy expressions for the universal functions $\mathcal{G}_{1,2}$ are given in Appendix \ref{appendix:G1G2}, an excellent uniformly valid solution can be constructed by simply adding the leading orders of the small and large $\lambda$ expansions of $\mathcal{G}_1$ (analogous to the function $\mathcal{F}$ in \cite{agarwal2021unrecognized}). Taylor expansion in both the viscously dominated limit ($\lambda \to 0$) and the inviscid limit ($\lambda \to \infty$) obtains
\begin{align}
		\mathcal{G}_1^{v} =-\frac{63}{80}\sqrt{\frac{3}{2\lambda}}+\mathcal{O}(1),\quad \mathcal{G}_1^{i}=-\frac{1}{2} + \mathcal{O}(1/\sqrt{\lambda})\,,
		\label{G1lamexp}
\end{align}
\begin{figure}
	\centering
	\includegraphics[width=\textwidth]{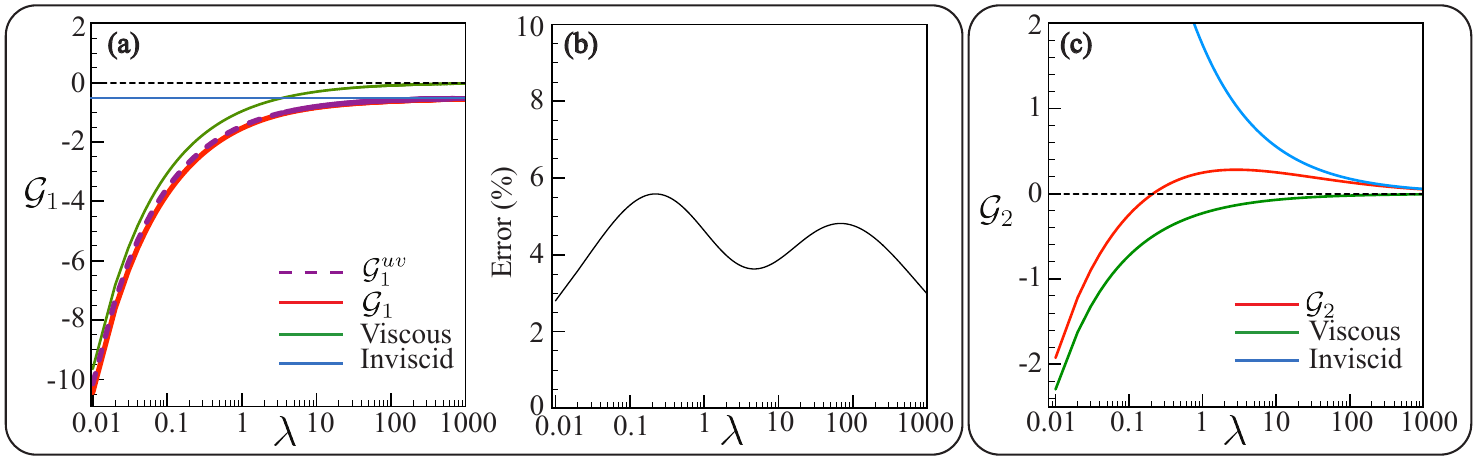}
	\caption{{\bf (a)} Plot of the in-phase force function $\mathcal{G}_1$. The uniformly valid expression (purple dashed) closely tracks the full solution (red). Also displayed are the viscous (green) and inviscid (blue) limit asymptotes. {\bf (b)} The magnitude of the percentage error between the uniformly valid and full solutions is small throughout the entire range of $\lambda$, with a maximum error of $\sim 6\%$. {\bf (c)} Plot of the out-of-phase force function $\mathcal{G}_2$ (red) together with its viscous (green) and inviscid (blue) limit expressions.} \label{fig:G1G2}
\end{figure}
from which the following uniformly valid result is constructed: 
\begin{align}
    \mathcal{G}_1^{uv}(\lambda) \approx -\left(\frac{1}{2} +\frac{63}{80}\sqrt{\frac{3}{2\lambda}}\right).\label{calG1twoterm}
\end{align}
Figure~\ref{fig:G1G2}(a,b) illustrates that this simple two-term expression agrees very well with the full result \eqref{G1full} over the entire range of the parameter $\lambda$, with a maximum error of $\sim 6\%$. 

A Taylor expansion of the out-of-phase term $\mathcal{G}_2$, in the viscous and inviscid limits, respectively, results in
\begin{align}
    \mathcal{G}_2^v = \frac{3}{16}\sqrt{\frac{3}{2\lambda}} +\mathcal{O}(1),\quad \mathcal{G}_2^i = -\frac{57}{40}\sqrt{\frac{3}{2\lambda}} + \mathcal{O}(1/\lambda).
\end{align}
Both of the above expansions have a $\O(1/\sqrt{\lambda})$ leading-order term and a simple, two-term approximation fails. In the following, we use the full expression \eqref{calG2}, noting that the contribution from this term is small in most practical situations, i.e., when $\lambda \gtrsim 1$. 

\section{Equation of motion for a particle immersed in an oscillatory flow}\label{sec:eom}
We now collect all the force contributions from Eqs.~\eqref{eq:F0}, \eqref{eqn:F1}, \eqref{F11F11}, \eqref{Fsiggam}, and combine them with the results of \cite{maxey1983equation} and \cite{agarwal2021unrecognized}. We use dimensional variables for easier physical interpretation. 
The following is the equation of motion for the velocity $\bm{U}_p$ of a rigid spherical particle immersed in an oscillatory background flow field $\bm{U}$, taking into account all force terms up to $\O(\operatorname{Re}_p)$:
\begin{subequations}
\begin{align}
    m_p \frac{d \bm{U}_p}{dt} =& \bm{F}^{(0)}_0+\bm{F}^{(1)}_0+ \operatorname{Re}_p\left(\bm{F}^{(0)}_1+\bm{F}^{(1)}_1\right)+O(\operatorname{Re}_p^2),\\
    \bm{F}^{(0)}_0=& m_f \frac{\partial \bm{U}}{\partial t}, \quad \operatorname{Re}_p\bm{F}^{(0)}_1= m_f \left(\bm{U}\cdot\nabla \bm{U}\right) + m_f a_p^2 \nabla\bm{U}:\nabla\left(\nabla \bm{U}\right)\mathcal{F}^{(0)}_1,\label{eqn:F0dim}\\
    \bm{F}^{(1)}_0=&- \frac{1}{2}m_f \frac{d}{dt}\left[\bm{U}_p-\bm{U}\right] - 6\pi \rho_f \nu a_p \left[\bm{U}_p(t) - \bm{U}(\bm{r}_p(t),t) \right] \nonumber\\
    &- 6\pi^{1/2} \nu^{1/2} a_p^2 \rho_f \int_{-\infty}^t \frac{d/d\tau \left[\bm{U}_p(t) - \bm{U}(\bm{r}_p(t),t)\right]}{\sqrt{t-\tau}}d\tau,\label{eqn:F10dim} \\
    \operatorname{Re}_p\bm{F}^{(1)}_1=& m_f \left[ (\bm{U}_p -\bm{U})\cdot \nabla \bm{U}\right]\, \mathcal{G}_1(\lambda) + m_f \left[ \partial_t(\bm{U}_p -\bm{U}) \cdot \nabla \bm{U}\right]\, \frac{\mathcal{G}_2(\lambda)}{\omega} \nonumber\\
    &+ m_f a_p^2 \nabla\bm{U}:\nabla\left(\nabla \bm{U}\right)\mathcal{F}^{(1)}_1.\label{eqn:F11dim}
\end{align}\label{eom_full_dim}\noindent
\end{subequations}
Here, we have dropped the contraction with $\bm{e}$ in \eqref{F11F11} and \eqref{Fsiggam} to derive $\bm{F}_0^{(1)}$, since the direction $\bm{e}$ is arbitrary (cf.\ the equivalent argument in \cite{maxey1983equation}). 
Equation~\eqref{eqn:F0dim} includes the background flow force term missing from MR mentioned in Sec.~\ref{sec:backgndforces}, proportional to $\mathcal{F}^{(0)}_1=1/5$.
Note that the scales of all the inviscid and inertial force terms use $m_f$,
while the viscous force terms contain $\nu$ explicitly.
In the following, we point out that \eqref{eom_full_dim}, while containing new physics, encompasses a number of earlier results as special cases, clarifying  connections between them. 

\subsection{Generalized Auton correction}\label{sec:genauton}
We first comment on the limiting case of the well-known correction to MR due to \cite{auton1988force}. The equation of motion derived by MR deviated from previous versions in the form of the convective term in \eqref{eqn:F0dim}, using $m_f \left(\bm{U}\cdot\nabla \bm{U}\right)$ instead of $m_f \left(\bm{U}_p\cdot\nabla \bm{U}\right)$ --- the values of these two derivatives can differ substantially when the Reynolds number is not small. Similarly, \citet{auton1988force} showed that in the limit of potential flows, the added mass term  should read $\frac{1}{2}m_f \left(\frac{d\bm{U}_p}{dt}-\frac{D\bm{U}}{Dt}\right)$ instead of $\frac{1}{2}m_f \left(\frac{d\bm{U}_p}{dt}-\frac{d\bm{U}}{dt}\right)$.
 Again, these two expressions are identical in the zero Reynolds number limit employed by MR, but in flows with substantial inertial effects, they can differ significantly.

Our formalism naturally addresses these concerns through the rigorous treatment of the disturbance flow around the particle. The first term on the RHS of \eqref{eqn:F11dim}, involving $\mathcal{G}_1$, modifies the added mass term in \eqref{eqn:F10dim} and reproduces the Auton correction \citep{auton1988force} in the inviscid, potential flow limit ($\lambda\to \infty$, $\operatorname{Re}_p\ll 1$), modifying $\frac{d \bm{U}}{dt}$ to $\frac{D \bm{U}}{Dt}$, or explicitly: 
\begin{align}
    &-\frac{1}{2}m_f \frac{d}{dt}\left[\bm{U}_p-\bm{U}\right]
    + m_f \left[ (\bm{U}_p -\bm{U})\cdot \nabla \bm{U}\right]
    \, \mathcal{G}_1(\lambda)\nonumber\\
    &\approx -\frac{1}{2}m_f \left[\frac{d}{dt}\bm{U}_p-\frac{D}{Dt}\bm{U}\right]
    - \frac{63}{80}\sqrt{\frac{3}{2\lambda}} m_f \left[ (\bm{U}_p -\bm{U})\cdot \nabla \bm{U}\right]\,,
     \label{eqn:G1}
\end{align}
where we use the simple two-term approximation \eqref{calG1twoterm} for $\mathcal{G}_1$. Thus, instead of heuristically modifying the added mass term, our approach rigorously derives its dependence on $\lambda$. Note that in most practically relevant oscillatory microfluidic flows, the value of $\lambda$ is ${\cal O}(1-10)$, so that the contribution from the second term of \eqref{eqn:G1}---capturing the effect of viscous streaming around the particle---results in the inertial force being quite large due to the $1/\sqrt{\lambda}$ scaling, reminiscent of Saffman lift \citep{saffman1965lift}. 

We note that the second term in \eqref{eqn:F11dim}, involving $\mathcal{G}_2$, arises due to the out-of-phase component of the slip velocity and thus characterises diffusion of vorticity from the particle.
This term is analogous to the Basset-Boussinesq history force and contributes most prominently when $\lambda\sim \mathcal{O}(1)$, while it is sub-dominant for both small and large $\lambda$. 

\subsection{Time-scale separation and connection to acoustofluidics}
Equation \eqref{eom_full_dim} describes unsteady particle dynamics as an integral equation containing a history integral, which can be explicitly evaluated in special cases, particularly  for particles executing purely oscillatory motion. In more general settings, where there is a superposition of \emph{slower} rectified or transport fluid flows---with a clear separation of scales from the \emph{fast} oscillatory motion---we can still find an explicit, analytical evaluation of the memory integral by employing the method of multiple scales. This approach results in a simple overdamped equation of motion for the particle that captures the slow dynamics accurately, as outlined in the following (see Appendix \ref{appendix:timescale} for details).

For flows induced by a localized oscillating source with curvature scale $a_b$, amplitude $\epsilon a_b$ and angular frequency $\omega$, we non-dimensionalize our equation with $a_b$, $\epsilon a_b \omega$ and $1/\omega$ as characteristic length, velocity, and time scales, respectively. Equation \eqref{eom_full_dim} then reads
\begin{align}
    \lambda \left( \hat{\kappa} + 1\right)\frac{d^2 \textbf{r}_p}{d t^2} =& \epsilon\lambda \frac{\partial \bm{u}}{\partial t} + \frac{2\lambda}{3} \epsilon^2 \bm{u}\cdot \nabla \bm{u} - \frac{\lambda}{3}\epsilon^2 \lambda \bm{U}_p \cdot \nabla \bm{u} -\left(\frac{d \textbf{r}_p}{d t}-\epsilon\bm{u}\right)\nonumber\\
    &+ \sqrt{\frac{3\lambda}{\pi}}\int_{-\infty}^t \frac{d/d\tau \left[d\bm{r}_p(\tau)/d\tau - \epsilon\bm{u}(\bm{r}_p(\tau),\tau)\right]}{\sqrt{t-\tau}}d\tau\nonumber\\
    & + \frac{2\lambda}{3}\epsilon\,\mathcal{G}_1  \left(\frac{d \textbf{r}_p}{d t}-\epsilon\bm{u}\right)\cdot \nabla \bm{u} + \frac{2\lambda}{3}\epsilon\, \mathcal{G}_2 \partial_t \left(\frac{d \textbf{r}_p}{d t}-\epsilon\bm{u}\right)\cdot \nabla \bm{u} \nonumber\\
    &+ \frac{2\lambda}{3}\epsilon^2 \alpha^2 \, \mathcal{F} \nabla\bm{u}:\nabla \nabla \bm{u},\label{eqn:eomunsteady}
\end{align}
where $\hat{\kappa}=2/3\left(\frac{\rho_p}{\rho_f}-1\right)$ is a dimensionless measure of density difference, $\alpha=a_p/a_b$ is the relative particle size, and  $\frac{d \bm{r}_p}{dt} =\epsilon \bm{u}_p$. As in \cite{agarwal2021unrecognized}, we write $\mathcal{F}=\mathcal{F}^{(0)}_1+\mathcal{F}^{(1)}_1$.

We employ standard techniques of time-scale separation (see Appendix \ref{appendix:timescale}) to obtain the leading order overdamped equation of particle motion. Briefly, the fast oscillatory dynamics in \eqref{eqn:eomunsteady} are time-averaged over the oscillation period and the resulting equation describes the dynamics of the leading-order mean particle position $\bm{r}_{p_0}$ on the slow time scale $T=\epsilon^2 t$,
\begin{align}
\frac{d \bm{r}_{p_0}}{d T}=\frac{\hat{\kappa}\lambda}{(\hat{\kappa}+1)} \mathcal{G}(\lambda) \left\langle \bm{u} \cdot \nabla \bm{u}\right\rangle + \frac{2\lambda}{3}\alpha^2 \mathcal{F}(\lambda) \left\langle \nabla \bm{u}:\nabla \nabla \bm{u} \right\rangle \,,\label{eom1dslowtime}
\end{align}
with $\mathcal{F}(\lambda)\approx\frac{1}{3} + \frac{9}{16}\sqrt{\frac{3}{2\lambda}}$ derived in \cite{agarwal2021unrecognized} and
\begin{align}
    \mathcal{G}(\lambda)&=\frac{(\hat{\kappa}+1)(2(1-\mathcal{G}_1)(d+\hat{\kappa})\lambda^2 +c\left(2\lambda\mathcal{G}_2-3\right))}{3(c^2+(d+\hat{\kappa})^2\lambda^2)}\,,\label{Gcal}
\end{align} 
where $c = 1+\sqrt{3\lambda/2}$ and $d=1+\sqrt{3/(2\lambda)}$ are expressions resulting from the integration of the history force term (cf.\ Appendix~\ref{appendix:timescale} for details). 

The first term on the RHS of \eqref{eom1dslowtime} can be rewritten as $F_{R}\mathcal{G}(\lambda)$, where $F_{R} = \frac{\hat{\kappa}\lambda}{(\hat{\kappa}+1)} \left\langle \bm{u} \cdot \nabla \bm{u}\right\rangle$ is a time-averaged force formally identical to the acoustic radiation force induced by an incident sound field with velocity field $\bm{u}$ \citep{bruus2012acoustofluidics}. In the acoustofluidic context, this velocity field may be caused by an oscillating object (bubble) excited by a primary acoustic wave. The resulting force from the bubble on a distant particle is then often denoted as
the secondary radiation force \citep{doinikov1996interaction}. 
As the acoustic formalism is based on the assumption of inviscid flow, $\mathcal{G}(\lambda)$ generalizes the far-field inviscid $F_{R}$ to include viscous effects that, as shown below, can change the resulting particle motion quantitatively and qualitatively. Note that ${\cal G}(\lambda\to\infty)=1$, recovering the inviscid case, while the viscous limit depends on the density contrast, ${\cal G}(\lambda\to 0)=-(1+\hat{\kappa})$.

In the next section, we specialize \eqref{eom1dslowtime} to the simplest case of a background flow induced by a volumetrically oscillating object---a situation commonly encountered in many practical microfluidic setups involving acoustically excited microbubbles---and compare our results with direct numerical simulations.

\section{Validation with Direct Numerical Simulations}\label{sec:results}
We have shown that the present analytical formalism generalizes previous attempts at predicting the behavior of particles in oscillatory flows. It is crucial to confirm the quantitative accuracy of our model. To this end, we compare our analytical predictions with independent, first-principles Direct Numerical Simulations (DNS) of the full Navier--Stokes equations, previously validated in a variety of streaming flow scenarios (see Refs.~ \cite{gazzola2011simulations,parthasarathy2019streaming,bhosale_parthasarathy_gazzola_2020,bhosale2022multicurvature,chan2022three,bhosale2022soft,pyaxisymflow2023} for details) and capturing the full dynamics of the fluid-particle system.

In order to make quantitative comparisons, we restrict the background flow field to a spherical, oscillatory monopole. These flows are typically generated near volumetrically excited bubbles and have been shown to actuate inertial forces on particles in oscillatory microfluidics \citep{rogers2011selective,chen2014manipulation,zhang2021portable}, showcasing their practical utility. This specialization offers an ideal framework for validating our analytical formalism, as this radially symmetric flow by itself does not induce viscous streaming, enabling us to neatly isolate the effect of inertial forces.

Accordingly, we insert 
$\bm{u}(r,t)= (1/r^2) e^{i t}\bm{e}_r$ into
eq.~\eqref{eom1dslowtime} to obtain the following time-averaged equation of motion (we drop the subscript $0$):
\begin{align}
    \frac{d r_p}{d T} &= -\frac{\hat{\kappa}\lambda}{r_p^5(\hat{\kappa}+1)}  \mathcal{G}(\lambda) -\frac{6}{r_p^7}\alpha^2 \lambda \mathcal{F}(\lambda),\label{eom1dslowtimemonopole}
    \end{align}
where $-\frac{\hat{\kappa}\lambda}{r_p^5(\hat{\kappa}+1)}=F_{R}$ and $r_p$ is in units of the radius of the oscillating source. This simple ODE provides clear predictions for the particle fate that can be compared with results from DNS. Two terms in \eqref{eom1dslowtimemonopole} determine the direction of  particle motion: The second term involving $\mathcal{F}$ is always negative \citep{agarwal2021unrecognized}, representing attraction towards the source while the sign of the first term changes with  $\hat{\kappa}$ and $\mathcal{G}$. Therefore, the magnitude and sign of the net force depend on several parameters, including $\lambda$, $\hat{\kappa}$, and also on $r_p$, as the first term dominates the second at large distances. 

Note that this setup is specifically constructed such that all effects on the RHS of \eqref{eom1dslowtimemonopole} are due to inertia. Thus, the comparison between  analytical predictions and DNS solutions provides a direct and accurate test of particle-inertial effects in oscillatory microfluidics. We will focus on the key quantities of practical interest: the particle trajectories, velocities and forces. 

\begin{figure}
	\centering
	\includegraphics[width=\textwidth]{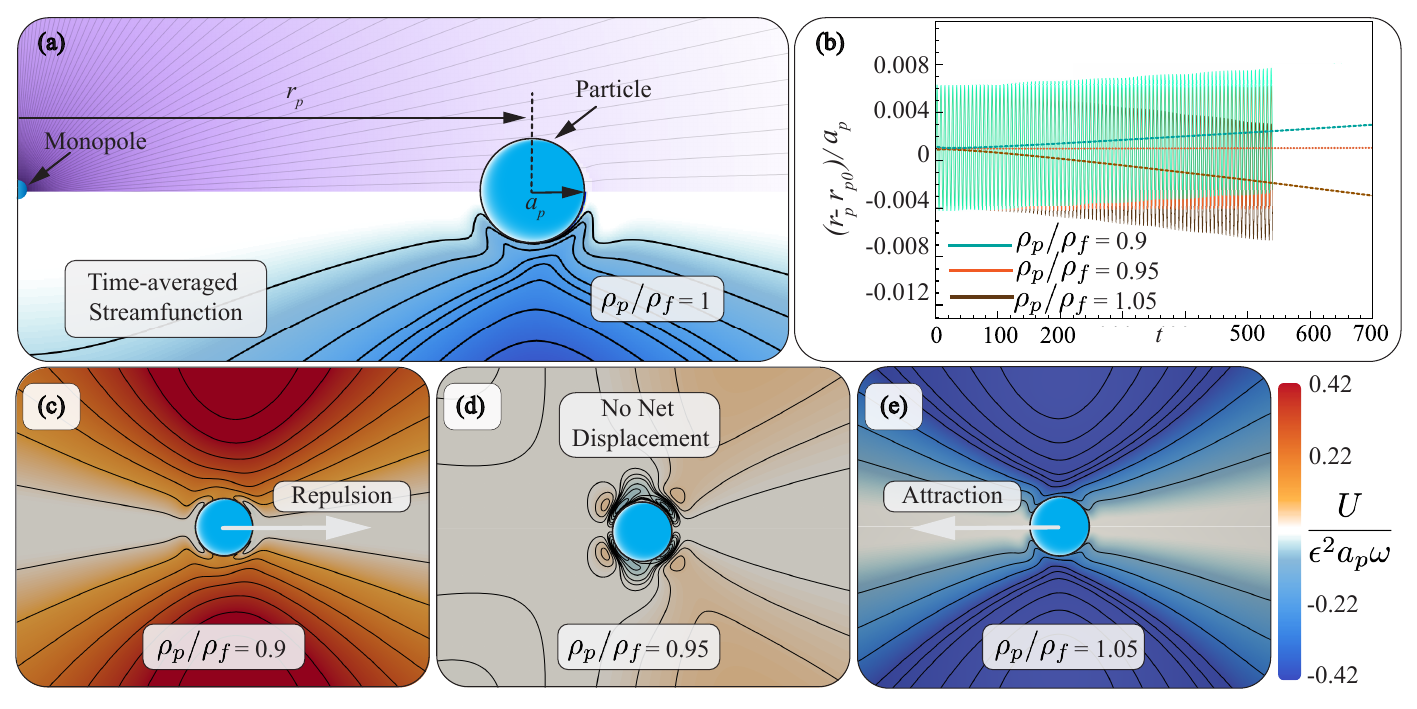}
	\caption{Direct numerical simulation of the prototypical problem: {\bf (a)} a spherical particle of radius $a_p$ is exposed to an oscillating monopole placed at a distance $r_p$ from the particle center with primary flow velocity $U^*$.
	Top figure: instantaneous streamlines (color bar is flow speed in units of $U^*$); bottom figure: time-averaged streamlines (color bar is steady flow speed in units of $\epsilon U^*$). {\bf (b)}  Particle coordinate as a function of time.
 For three different density contrasts, we show the full oscillatory dynamics as well as the steady particle motion (averaged once per oscillation cycle). {\bf (c-e)}: Time-averaged flow fields around the particle for the three cases of (b).} \label{fig:DNSresults}
\end{figure}
\subsection{Simulation approach and results}
To computationally simulate the relevant flow scenarios, we employ an axisymmetric formulation of the incompressible Navier Stokes equations (see Appendix \ref{appendix:DNS}). 
Figure~\ref{fig:DNSresults}a presents the simulation set-up. A spherical particle of radius $a_p$ is initially released with zero velocity at a distance $r_{p0}$ from the oscillating monopole. It is thus exposed to the model flow of frequency $\omega$ and velocity amplitude $\epsilon\omega$ (the nominal source size $a_b$ is normalized to 1). We choose $\epsilon=0.01$, $a_p=0.05$, $\omega = 16\pi$ throughout, and $r_{p0}=2$ unless otherwise stated. The fluid viscosity is determined from the corresponding values of $\lambda$ in each simulation. The upper half of Fig.~\ref{fig:DNSresults}a shows representative streamlines of the instantaneous, near-radial flow, while the bottom half shows time-averaged streamlines, highlighting the ensuing steady, rectified flow pattern. Varying the ratio of particle density to fluid density in Fig.~\ref{fig:DNSresults}(c,d, and e) while keeping all other parameters constant shows that the direction of this rectified flow reverses, but not for matching densities---rather, the flow pattern loses directionality around $\rho_p/\rho_f\approx 0.95$.

Accordingly, the particle motion in the simulation (Fig.~\ref{fig:DNSresults}b) reverses direction: particles lighter than $\approx 0.95\rho_f$ are repelled over time, while those of greater density are attracted towards the monopole (which includes the density-matched case). 

\begin{figure}
	\centering
	\includegraphics[width=\textwidth]{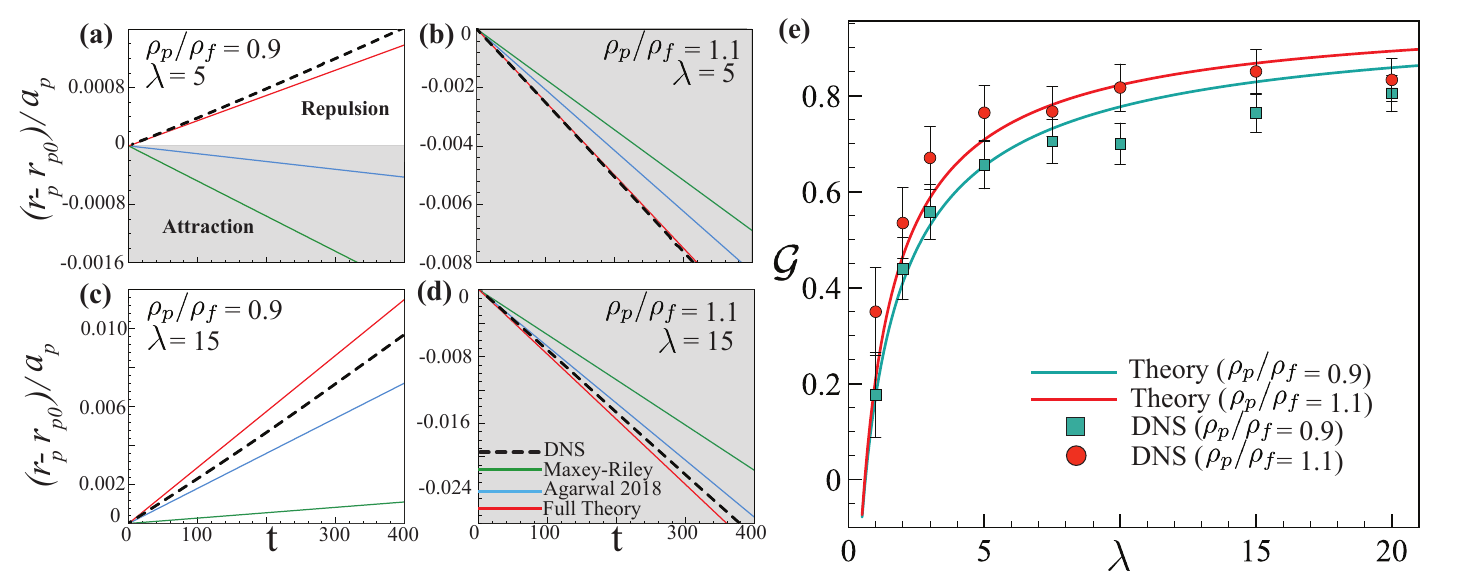}
		\caption{ 
  Comparison of theoretical particle motion with DNS.
  {\bf (a-d)} Time-averaged dynamics from the  theory using Eq.~\eqref{eom1dslowtimemonopole} with the full analytical expressions for $\mathcal{G}$ and $\mathcal{F}$ agree with DNS (magenta) for the entire range of $\lambda$ and density contrast values (all results are for $r_{p0}=2$). Two exemplary density contrast and $\lambda$ combinations are displayed, $\rho_p/\rho_f=1.1$ ($\hat{\kappa}=0.067)$ and $\rho_p/\rho_f=0.9$ ($\hat{\kappa}=-0.067)$. The classical MR equation solutions (green) fail to even qualitatively capture the particle repulsion in (a), and otherwise strongly underestimate the  force (b-d). The inviscid formalism of \cite{agarwal2018inertial} (light blue) has similar, though quantitatively less severe, shortcomings. {\bf (e)} Best-fits of $\mathcal{G}(\lambda)$ to \eqref{eom1dslowtimemonopole} are extracted from DNS and show excellent agreement with the full theory \eqref{Gcal}, for both heavier ($\rho_p/\rho_f=1.1$, red) and lighter ($\rho_p/\rho_f=0.9$, teal) particles.} \label{fig:traj_Gcal}
\end{figure}

\subsection{Comparison of particle trajectories}

A comparison between unsteady DNS dynamics and predictions from the unsteady theory equation \eqref{eqn:eomunsteady} is possible, although it entails evaluating the non-local Basset memory integral, which is computationally expensive and typically not of relevance in applications.
For a clearer and more practical validation, in Fig.~\ref{fig:traj_Gcal} we focus  on comparing time-averaged DNS dynamics and predictions from the analytically derived  equation \eqref{eom1dslowtimemonopole} for the rectified steady dynamics, which is easily integrated in time. 

Figures \ref{fig:traj_Gcal}(a-d) depict examples of such averaged radial dynamics for different density ratios and different Stokes numbers $\lambda$, all with $r_{p0}=2$. Across a wide range of parameters,  DNS dynamics (magenta) and  analytical results (red) from the uniformly valid asymptotic expressions of ${\cal G}(\lambda)$ are found to be in very good agreement. Predictions from the classical MR equation (green) instead deviate significantly in all cases and, for some parameter combinations (see Fig.~\ref{fig:traj_Gcal}a), even misidentify the direction of the particle motion. 
The theory of \cite{agarwal2018inertial} (light blue), which relies on inviscid flow throughout, also misses important force contributions and shows deviations similar in nature to those of MR, though quantitatively smaller. Only properly accounting for particle inertia successfully reproduces the range of numerically observed behaviors.

To illustrate the success of Eq.~\eqref{eom1dslowtimemonopole} over the entire range of practically relevant $\lambda$ values, Fig.~\ref{fig:traj_Gcal}(e) condenses all results by extracting a ${\cal G}(\lambda)$ value from 
best-fitting \eqref{eom1dslowtimemonopole} to
the numerically simulated particle trajectories (see Appendix \ref{appendix:fit} for details), given the previously established accuracy of the ${\cal F}(\lambda)$ function \citep{agarwal2021unrecognized}. Both for heavier 
($\rho_p/\rho_f=1.1$, red) and lighter particles ($\rho_p/\rho_f=0.9$, teal), the analytical equation yields excellent agreement with the simulated rectified drift of the particle, indicating that it captures the key physical mechanisms at play. We note here that even for $\lambda=20$, there are significant deviations of ${\cal G}(\lambda)$ from its inviscid asymptotic value of 1, showing that viscous effects remain important in quantitative device design even at large Stokes numbers. 
 
This validation demonstrates the utility of our theoretical framework in predicting the dynamics of solid particles in oscillatory flows, as each individual DNS simulation incurs a large computational cost up to $\sim$24-48 core hours on a single node on the Expanse supercomputer (see Appendix~\ref{appendix:DNS}), while the theory ODE is trivial to solve.


\subsection{Particles at large distances: Connection to Acoustofluidics}
\begin{figure}
		\centering
		\includegraphics[width=\textwidth]{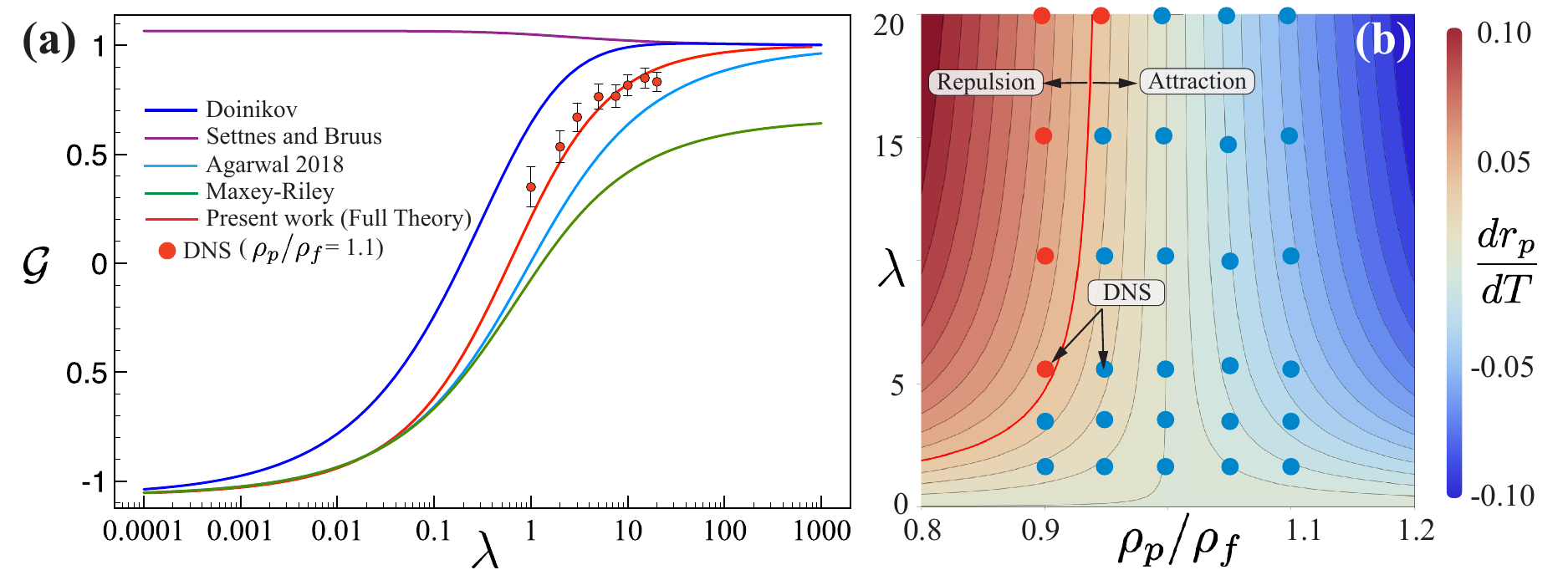}
		\caption{{\bf (a)} Stokes number dependence of the overall dimensionless inertial force magnitude ${\mathcal G}$, representing the ratio between acoustofluidic forces (limit of large distance between source and particle) to the radiation force $F_R$. Lines are results from different theories, symbols from DNS, all for $\rho_p/\rho_f$ = 1.1 ($\hat{\kappa}=0.067$), $\epsilon=0.01$. The DNS values are best fits of ${\cal G}$ given the full expression for ${\cal F}$ in \eqref{eom1dslowtimemonopole}. The present work (red line) is in excellent agreement with all DNS data, while both the \cite{agarwal2018inertial} (light blue) and Maxey--Riley formalisms (green) significantly underestimate the forces. {\bf (b)} Contour plots for steady particle velocity at $r_p=2$ with varying $\lambda$ and  $\rho_p/\rho_f$. The solid red line marks the transition from attraction to repulsion.  Solid circles indicate simulation outcomes with blue and red circles representing attraction and repulsion, respectively. } \label{fig:Gcal_sim_theory}
	\end{figure}

Acoustofluidics has been a fruitful field of study aiming to manipulate fluid and particles using acoustic waves. As mentioned above, our framework specializes to the far-field acoustofluidic secondary radiation force when the distance between the particle and the oscillating source is large, $r_{p_0}\gg 1$. In this case, the force on the particle is the first term of equation \eqref{eom1dslowtime}, i.e., the nominal inviscid acoustic radiation force $F_{R}$ multiplied by the Stokes number-dependent factor ${\cal G}(\lambda)$. That such a $\lambda$-dependence exists has been known in acoustofluidics, and several approaches have been used to quantify it. We compile these predictions in Fig.~\ref{fig:Gcal_sim_theory}(a) for reference.

Predictions using the MR equation \citep{maxey1983equation} fail to correctly reproduce the inviscid limit ($\lambda\to \infty$) due to the incorrect form of the fluid acceleration in the added mass term (see the discussion of the Auton correction in section~\ref{sec:genauton}). The formalism of \cite{settnes2012forces} instead misses the opposite viscous limit ($\lambda\to 0$), as it ignores viscosity completely.
In previous work \citep{agarwal2018inertial}, the present authors heuristically combined the leading-order inviscid and viscous effects. This simplified formalism agrees exactly with the much more elaborate theory of \cite{doinikov1994acoustic} in both the viscously-dominated ($\lambda\to 0$) and the inviscid limits ($\lambda\to \infty$), while quantitative discrepancies remain in the intermediate $\lambda$ regime, where the ${\cal G}(\lambda)$ of \cite{doinikov1994acoustic} is larger than that of \cite{agarwal2018inertial}.
 
The theory of the present work agrees with the previously established viscous and inviscid limits, and makes new predictions in the intermediate $\lambda$ range, with values in-between  those of \cite{doinikov1994acoustic} and \cite{agarwal2018inertial}. The DNS data in Fig.~\ref{fig:Gcal_sim_theory}a demonstrate that our theory is in excellent agreement with the forces observed in a full Navier-Stokes simulation, significantly improving on all previous approaches. The  relative error between our analytical predictions and the DNS is $\approx 5-10\%$ across the simulation range $1\leq \lambda\leq 20$.


Our results reaffirm that viscous effects can significantly affect the behavior of particles in acoustofluidic systems, and 
have important implications for the design and optimization of microfluidic devices that utilize acoustic waves for particle manipulation.

\subsection{Transition from attraction to repulsion}\label{sec:attrep}
Equation \eqref{eom1dslowtimemonopole} predicts that particles in monopolar oscillatory flows can exhibit equilibrium positions (at finite $r$) where the net force acting on the particle is zero. Setting $\frac{d r_p}{d T}=0$ 
in \eqref{eom1dslowtimemonopole} obtains the critical radial position (in units of the particle radius $a_p$) as
\begin{align}
    r_{p_c}= \sqrt{-\frac{(\hat{\kappa}+1)\mathcal{F}(\lambda)}{\hat{\kappa}\mathcal{G}(\lambda)}}\,.\label{eqn:rpc}
\end{align}
In most practically relevant situations, $\lambda\gtrsim \mathcal{O}(1)$, and thus  $\mathcal{G}>0$ (cf.\  Fig.~\ref{fig:Gcal_sim_theory}a). 
A real $r_{p_c}$ then exists 
if the particle is lighter than the surrounding medium ($\hat{\kappa}<0$). Such an equilibrium position is necessarily unstable, as the repulsive term in \eqref{eom1dslowtimemonopole} decays more slowly. Thus, 
for light particles and $\lambda\gtrsim \mathcal{O}(1)$ this model predicts a critical radial distance below which the particle is always attracted towards the oscillating source. In a practical set-up, a particle can be transported into this attractive range by streaming flows or other appropriately designed flow fields. Thus, $r_{p_c}$ is an important quantity to consider in the design of microfluidic devices that make use of acoustically excited microbubbles to selectively trap particles (cf. \citet{chen2016onset,zhang2021portable,zhang2021versatile}).

Conversely, given a certain distance from the oscillating object, attraction or repulsion of a particle can be designed by  adjusting density contrast or Stokes number (oscillation frequency). 
Figure~\ref{fig:Gcal_sim_theory}(b) plots the iso-lines of the RHS of \eqref{eom1dslowtimemonopole} as a function of the parameters $\lambda$ and $\hat{\kappa}$, for a fixed $r_{p}=2$. The red line is the zero contour separating attractive from repulsive dynamics. Particles of density equal or higher than fluid density are always attracted towards the source, while light paticles ($\rho_p/\rho_f<1$) are repelled above a threshold Stokes number. Comparison with DNS data (circles in Fig.~\ref{fig:Gcal_sim_theory}(b)) confirms these predictions.
The sign change of ${\cal G}(\lambda)$ at $\lambda\ll 1$ complicates this picture (in principle, repulsion can be achieved even for heavier particles), although  force magnitudes in this regime are typically too small to be practically relevant.

\section{Relevance and limitations of the inertial equation of motion}\label{sec:disc}
\subsection{Avoiding effects of outer-flow inertia}\label{sec:outerflow}
The results obtained in this study show that particle motion can be described quantitatively by inertial forcing terms. Often, such computations are complicated by a transition between a viscous-dominated inner flow volume (near the particle) and an inertia-dominated outer volume, necessitating an asymptotic matching of the two limits, such as for the Oseen \citep{oseen1910uber} and Saffman \citep{saffman1965lift} problems. Our formalism, however, only employs an inner-solution expansion and still obtains accurate predictions. This can be rationalized by invoking the analysis of \cite{lovalenti1993hydrodynamic}, who showed that an outer region is not present when the magnitude of oscillatory inertia in the disturbance flow $\partial \boldsymbol{w}^{(1)}/\partial t$ is much larger than that of the advective term ${\bm f}$, i.e., the characteristic unsteady time scale $\omega^{-1}$ is shorter than the convective inertial time scale $\nu/(U^*w^{(0)})^2$, where $w^{(0)}$ is the dimensionless velocity scale of the fluid in the particle reference frame. For the case of non-neutrally buoyant particles, $w^{(0)} = {\cal O}(\hat{\kappa})$, so that the criterion becomes 
\begin{align}
\epsilon^2 \lambda \ll \min(\alpha^2/\hat{\kappa}^2,1).\label{eqn:crit}
\end{align}
As long as the density contrast between the particle and fluid is small, or $|\hat{\kappa}|\ll 1$, \eqref{eqn:crit} is easily satisfied in most experimental situations, and it reverts to the criterion $\epsilon^2\lambda \ll 1$ established for neutrally buoyant particles \citep{agarwal2021unrecognized}. An interesting point to note is that the density-dependent condition  $\epsilon^2 \lambda \ll \alpha^2/\hat{\kappa}^2$ can be rewritten in the $a_p$-independent form $\epsilon\ll \delta_S/(a_b\hat{\kappa})$. This is because the leading term of the background flow field expansion at the particle  position contains no information about the particle length scale. 

\subsection{Magnitude and practical relevance of inertial effects}\label{sec:effects}
In Fig.~\ref{fig:Gcal_sim_theory}a, we illustrated how our formalism, in agreement with DNS, predicts much stronger inertial forces than either \citet{maxey1983equation} (which emphasizes viscous effects) or \citet{agarwal2018inertial}, which treats the background flow as inviscid. For particles typically encountered in microfluidic applications involving biological cells, the density difference tends to be around $5\%$, while the size parameter is $\alpha\lesssim 0.2$ and $\lambda \gtrsim 1$. A practically useful metric to quantify the effect of the inertial force acting on the particle by a localized oscillating source is the time needed for radial displacement of a particle diameter. In most particle manipulation strategies, $r_p\gtrsim a_b$ and, upon solving \eqref{eom1dslowtimemonopole} with these nominal parameter values, we find that our formalism predicts a timescale of $\sim 10$ms compared to $\sim 50$ms predicted by the inviscid formalism. This translates to much more efficient design strategies for sorting particles based on size or density. The MR formalism predicts a time scale of $\sim 500$ms, which is off by more than one order of magnitude and severely underestimates the performance of oscillatory microfluidic set-ups. 

For these prototypical cases where particles are close to the interface of the oscillating object ($r_p\gtrsim a_b$), the major contribution to inertial forces is due to $F_{\Gamma \kappa}$,
as discussed in \cite{agarwal2021unrecognized}, However, since $F_{\Gamma \kappa}$ decays more strongly with the distance from the source than $F_{\sigma\Gamma}$, the density contrast dependent force can easily become comparable in magnitude, resulting in the rich behavior of attraction and repulsion separated by the critical (and tunable) distance $r_{p_c}$ as described in Sec.~\ref{sec:attrep}. Thus, present work suggests new avenues for particle trapping/sorting relying on density contrast; some of these ideas will be explored in future publications.

In a microfluidic set-up, the oscillatory flow is induced around an obstacle, e.g.\ a cylinder or bubble of radius $a_b$, and as mentioned above, particles in typical applications will approach quite closely to the interface of this obstacle.
We have not accounted for effects due to such a nearby boundary in this analysis. In \cite{agarwal2018inertial}, we demonstrated the existence of a stable fixed point position when the particle is in very close proximity to the boundary. This stable equilibrium is a consequence of the repulsive lubrication force near the interface balancing the attractive force discussed here. As long as $|\hat{\kappa}|\ll 1$, which is the case in an overwhelming majority of practical applications, the conclusions of \cite{agarwal2018inertial} are not affected by the present findings, i.e., a particle attracted to an oscillating obstacle is expected to come to rest at a stable equilibrium distance that is extremely small compared with the interface scale, and typically even compared with the particle scale.

\section{Conclusions}\label{sec:concl}
We have developed a rigorous formalism to accurately describe the motion of particles in general, fast oscillatory flows. The present work systematically accounts for finite inertial forces in viscous flows that result from the interaction between the density-contrast dependent slip velocity and flow gradients. Confirmed by direct numerical simulations, these forces are found to be important and often far larger than the density-contrast dependent effects present in the original Maxey-Riley formalism. 
Our theory allows for quantitative predictions of the sign and magnitude of forces exerted on particles in 
many customary microfluidic settings, in particular for nearly density matched cell-sized particles---the most relevant case in medicine and health contexts.
The theory encompasses special cases such as Auton's correction and acoustic radiation forces in the inviscid limit, and provides their quantitative generalization in the presence of viscous effects.

\vspace{0.3cm}

{\em Acknowledgments:} The authors thank Bhargav Rallabandi and Howard Stone for helpful discussions. G.U. and M.G. acknowledge support under NSF CAREER $\#$1846752. Declaration of Interests: The authors report no conflict of interest.



\appendix

\section{Leading-order disturbance flow and mobility tensors}\label{appendix:mobility}
The leading-order equations for ($\boldsymbol{w}^{(1)}_0$, $p^{(1)}_0$) are unsteady Stokes and read
\begin{subequations}
	\begin{align}
			\nabla^{2} \boldsymbol{w}^{(1)}_0-\nabla p^{(1)}_0 &=3\lambda \frac{\partial \boldsymbol{w}^{(1)}_0}{\partial t}, \\
			\boldsymbol{\nabla} \cdot \boldsymbol{w}^{(1)}_0 &=0, \\
			\boldsymbol{w}^{(1)}_0 &= \bm{u}_{p_0}-\bm{u} \quad \text { on } \bm{r}=1, \label{eqn:w10bc}\\
			\boldsymbol{w}^{(1)}_0 &= 0 \quad \text { as } \bm{r} \rightarrow \infty.
	\end{align}\label{eqn:w01}\noindent
\end{subequations} 
As a consequence of \eqref{eqn:uexp}, the boundary condition \eqref{eqn:w10bc} is also expanded around  $\bm{r}_{p_0}$, so that in the particle-fixed coordinate system
\begin{align}
	     \boldsymbol{w}^{(1)}_0 = \bm{u}_{p_0}-\bm{u}  = \bm{u}_{p_0} -\bm{u}|_{\bm{r}_{p_0}} - \bm{r}\cdot \bm{E} - \bm{r}\bm{r}:\bm{G}+ \dots \quad \text{on} \quad\bm{r}=1\,, \label{w10genbc}
\end{align}
where we have retained the first three terms in the background flow velocity expansion. Owing to the linearity of the leading order unsteady Stokes equation, the solution can generally be expressed as \citep{landau1959course,pozrikidis1992boundary}
	\begin{align}
	\boldsymbol{w}^{(1)}_0  =  \bm{\mathcal{M}}_D \cdot \bm{u}_s  - \bm{\mathcal{M}}_Q \cdot\left(\bm{r}\cdot \bm{E}\right) -\bm{\mathcal{M}}_O \cdot\left(\bm{r}\bm{r}:\bm{G}\right) + \dots, \label{app:w10gen}
\end{align}
where $\bm{\mathcal{M}}_{D,Q,O}(r,\lambda)$ are spatially dependent mobility tensors. 

For harmonically oscillating, axisymmetric background flows (i.e., $\bm{u}(\bm{r},t)=\{\bar{u}_r,\bar{u}_\theta,0\} e^{i t}$ in the spherical particle coordinate system), general explicit expressions can be derived for the mobility tensors $\bm{\mathcal{M}}_{D,Q,O}$, ensuring no-slip boundary conditions on the sphere order-by-order. A procedure obtaining $\bm{\mathcal{M}}_D$ is described in \cite{landau1959course}; the other tensors are determined analogously. Using components in spherical coordinates, they read
	\begin{align}
		\bm{\mathcal{M}}_D = \begin{bmatrix}
			\frac{2a(r)}{r^2}  & 0 & 0 \\
			0 & \frac{a'(r)}{r} & 0 \\
			0 & 0 & 0
		\end{bmatrix}, \quad
		\bm{\mathcal{M}}_Q = \begin{bmatrix}
			\frac{b(r)}{r^3}  & 0 & 0 \\
			0 & \frac{b'(r)}{3r^2} & 0 \\
			0 & 0 & 0
		\end{bmatrix}, \quad
		\bm{\mathcal{M}}_O = \begin{bmatrix}
			\frac{-32c(r)}{3r^4}  & 0 & 0 \\
			0 & \frac{8c'(r)}{3r^3} & 0 \\
			0 & 0 & 0
		\end{bmatrix},
	\end{align}
	where
	\begin{subequations}
		\begin{align}
			a(r)=&\frac{1}{2\beta^2 r} \left[\beta^2 - 3i\beta+3 - 3 e^{-i\beta(r-1)}\left(1+i\beta r\right) \right],\\
			b(r)=& \frac{1}{\beta^2 (\beta-i) r^2}\left[\beta  (-15+\beta  (\beta -6 i))+15 i +5 e^{-i \beta  (r-1)} (\beta  r (3+i \beta  r)-3 i)\right],\\
			c(r)=&\frac{-3 (105+\beta  (\beta  (-45+\beta  (\beta -10 i))+105 i))}{32 \beta ^2 (-3+\beta  (\beta -3 i)) r^3}\nonumber\\
			&+\frac{21 e^{-i \beta  (r-1)} (15+\beta  r (-\beta  r (6+i \beta  r)+15 i))}{32 \beta ^2 (-3+\beta  (\beta -3 i)) r^3},
		\end{align}
	\end{subequations}
and $\beta = \sqrt{-ia_p^2/(\nu/\omega)}=\sqrt{-3i\lambda}$ is the complex oscillatory boundary layer thickness. We emphasize that these expressions are the same for arbitrary axisymmetric oscillatory $\bm{u}$. Accordingly, only the expansion coefficients $\bm{u}_s $, $\bm{E}$, and $\bm{G}$ contain information about the particular flow.

It is understood everywhere that physical quantities are obtained by taking real parts of these complex functions.

\section{Reciprocal theorem and test flow}\label{appendix:reciprocal}
In order to compute the force $\bm{F}^{(1)}_1$, we do not solve for the flow field $\boldsymbol{w}^{(1)}_1$ but instead employ a reciprocal relation in the Laplace domain to directly obtain the force. A key simplification due to oscillatory flows is that the Laplace transforms are explicitly computed. The symmetry relation employs a known test flow (denoted by primed quantities such as $\bm{u}'$) in a chosen direction $\bm{e}$, around an oscillating sphere such that it satisfies the following unsteady Stokes equation:
\begin{subequations}
	\begin{align}
			\nabla^{2} \bm{u}'-\nabla p' &=\nabla \cdot \bm{\sigma}'=3\lambda\frac{\partial \bm{u}'}{\partial t}, \\
			\boldsymbol{\nabla} \cdot \bm{u}' &=0, \\
			\bm{u}' &= u'(t)\,\bm{e} \quad \text { on } \bm{r}=1, \\
			\bm{u}' &= 0 \quad \text { as } \bm{r} \rightarrow \infty,
	\end{align}\label{test}\nolinebreak
\end{subequations}
where the unit vector $\bm{e}$ is chosen to coincide with the direction in which the force on the particle is desired. The solution to this problem is of the same form as \eqref{w10gen}, but with only the first term, i.e.,
	\begin{align}
		\bm{u}' = u'(t) \bm{\mathcal{M}}_D \cdot \bm{e}\,.
	\end{align}
Denoting Laplace transformed quantities by hats (e.g., $\hat{\bm{u}}$), the following symmetry relation is obtained using the divergence theorem (cf. \cite{lovalenti1993hydrodynamic,maxey1983equation,hood2015inertial}):
\begin{align}
		\oint_S  (\hat{\boldsymbol{w}}^{(1)}_1\cdot \hat{\bm{\sigma}}' - \hat{\bm{u}}'\cdot \hat{\bm{\sigma}}^{(1)}_1)\cdot \bm{m} \, dS= \int_V \left[\nabla \cdot (\hat{\boldsymbol{w}}^{(1)}_1\cdot \hat{\bm{\sigma}}') - \nabla \cdot (\hat{\bm{u}}'\cdot \hat{\bm{\sigma}}^{(1)}_1)\right]dV,
\end{align}
where $\bm{m}$ is the outward unit normal vector to the surface (pointing inward over the sphere surface), and $\hat{\bm{\sigma}} = \nabla \hat{\bm{u}} +(\nabla \hat{\bm{u}})^T - \hat{p}\bm{I}$. Substituting boundary conditions from \eqref{orep} and \eqref{test}, and setting the volume equal to the fluid-filled domain, we obtain
\begin{align}
		&\hat{\bm{u}}_{p_1}^{(1)} \cdot \int_{S_p} ( \hat{\bm{\sigma}}'\cdot \bm{m})dS - \hat{u}' \bm{e} \cdot \int_{S_p} ( \hat{\bm{\sigma}}^{(1)}_1\cdot \bm{m})dS +  \int_{S_\infty} ( \hat{\boldsymbol{w}}^{(1)}_1 \cdot\hat{\bm{\sigma}}')\cdot \bm{m}dS \nonumber\\
		&- \int_{S_\infty} ( \hat{\bm{u}}' \cdot\hat{\bm{\sigma}}^{(1)}_1)\cdot \bm{m}dS \nonumber\\
		=& \int_V \left[\hat{\boldsymbol{w}}^{(1)}_1 \cdot (\nabla\cdot \hat{\bm{\sigma}}') - \hat{\bm{u}}' \cdot (\nabla\cdot \hat{\bm{\sigma}}^{(1)}_1) + \nabla\hat{\boldsymbol{w}}^{(1)}_1 :\hat{\bm{\sigma}}' - \nabla\hat{\bm{u}}' :\hat{\bm{\sigma}}^{(1)}_1 \right]dV\,.\label{eqn:reciprocalRep}
\end{align}
The third term on the LHS is $0$ since the viscous test flow stress tensor decays to zero at infinity. Similarly, the integral in the fourth term vanishes in the far field if viscous stresses dominate inertial terms, and also in the case of inviscid irrotational flows (see \cite{lovalenti1993hydrodynamic,stone2001inertial}). The third and fourth terms on the RHS also go to zero, owing to incompressibilty and symmetry of the stress tensor:
	\begin{align}
		&\nabla\hat{\boldsymbol{w}}^{(1)}_1 :\hat{\bm{\sigma}}' - \nabla\hat{\bm{u}}' :\hat{\bm{\sigma}}^{(1)}_1= \nabla\hat{\boldsymbol{w}}^{(1)}_1 :(\nabla\hat{\bm{u}}'+(\nabla\hat{\bm{u}}')^T)-\hat{p}' \nabla \cdot \hat{\boldsymbol{w}}^{(1)}_1\nonumber\\
		&- \nabla\hat{\bm{u}}' : (\nabla\hat{\boldsymbol{w}}^{(1)}_1+(\nabla\hat{\boldsymbol{w}}^{(1)}_1)^T)-\hat{p}^{(1)} \nabla \cdot \hat{\bm{u}}'=0\,.
	\end{align}
The divergence of the hatted stress tensors in the remaining two terms of the RHS of \eqref{eqn:reciprocalRep} can be obtained by taking the Laplace transforms of \eqref{orep} and \eqref{test} and using the property $\widehat{f'(t)} = s\widehat{f(t)}-f(0)$, so that
	\begin{subequations}
		\begin{align}
			\nabla\cdot \hat{\bm{\sigma}}' &= 3\lambda s \hat{\bm{u}}' - \bm{u}'(0),\\
			\nabla\cdot \hat{\bm{\sigma}}^{(1)}_1 &= 3\lambda s \hat{\boldsymbol{w}}^{(1)}_1 - \boldsymbol{w}^{(1)}_1(0) + \hat{\bm{f}}_0\,.
		\end{align}
	\end{subequations}
Now, the force on the sphere at this order is given by $\bm{F}^{(1)}_1 =\int_{S_p} ( \bm{\sigma}^{(1)}_1\cdot \bm{n})dS=-\int_{S_p} ( \bm{\sigma}^{(1)}_1\cdot \bm{m})dS$, since $\bm{m}$ points inwards while $\bm{n}$ points outwards on the surface of the sphere. Assuming both $\boldsymbol{w}^{(1)}_1$ and $\bm{u}'$ start from rest, \eqref{eqn:reciprocalRep} simplifies to (cf. \cite{lovalenti1993hydrodynamic})
	\begin{align}
		\hat{u}' \bm{e}\cdot \frac{\hat{\bm{F}}^{(1)}_1}{F_S/(6\pi)} = \hat{\bm{u}}_{p_1} \cdot \int_{S_p} ( \hat{\bm{\sigma}}'\cdot \bm{n})dS - \int_V \hat{\bm{u}}' \cdot \hat{\bm{f}}_0dV\,.\label{eqn:F11lap}
	\end{align}
Taking the inverse Laplace transform, we obtain the expression for $\bm{e}\cdot\bm{F}^{(1)}$ given in the main text.
\section{Evaluation of $\mathcal{G}_1$ and $\mathcal{G}_2$} \label{appendix:G1G2}
\allowdisplaybreaks
In order to get explicit results for the non-trivial integration factors $\mathcal{G}_1$ and $\mathcal{G}_2$, we insert $\hat{\bm{f}}_0$ (with explicitly known mobility tensors $\bm{\mathcal{M}}_{D,Q,O}$) into \eqref{F11F11}. Since $F_{\sigma \Gamma}^{(1)}$ involves products of oscillatory terms, there are higher-order force harmonics with zero net effect on the particle dynamics which we will average out in the following to simplify the integration evaluations. 

We first decompose the slip velocity into its in-phase and out-of-phase components, i.e., $\bm{u}_s(\bm{r}_p,t)=\bm{u}_{s}^{I}(\bm{r}_p,t)+\bm{u}_{s}^{O}(\bm{r}_p,t)$, as noted in the main text, and time-average \eqref{F11F11} over a period of oscillation to remove higher-order harmonic terms. We then perform the volume integration to obtain an explicit but rather lengthy expression that can be symbolically written as
\begin{align}
 	\frac{\langle F_{\sigma \Gamma}^{(1)}\rangle}{\operatorname{Re}_p F_S/(6\pi)}=\frac{4\pi}{3}\langle\bm{u}_{s}^{I}\cdot\bm{E}\rangle \cdot \bm{e}\, \mathcal{{G}}_1(\lambda) +  \frac{4\pi}{3}\langle\bm{u}_{s}^{O}\cdot\bm{E}\rangle \cdot \bm{e}\, \mathcal{{G}}_2(\lambda).
\end{align}
where $\mathcal{G}_1$ and $\mathcal{G}_2$ are explicit outcomes of the volume integration. Exploiting the orthogonality of trigonometric functions and the fact that, for fast oscillatory background flows, $\bm{E}$ is purely in-phase, we rewrite the in-phase component as $\langle \bm{u}_{s}\cdot\bm{E}\rangle$ and the out-of-phase component as $\langle \partial_t \bm{u}_{s}\cdot\bm{E}\rangle$, where angled brackets denote time-averaging, so that
\begin{align}
 	\frac{\langle F_{\sigma \Gamma}^{(1)}\rangle}{\operatorname{Re}_p F_S/(6\pi)}=\frac{4\pi}{3}\langle\bm{u}_{s}\cdot\bm{E}\rangle \cdot \bm{e}\, \mathcal{{G}}_1(\lambda) +  \frac{4\pi}{3}\langle\partial_t \bm{u}_{s}\cdot\bm{E}\rangle \cdot \bm{e}\, \mathcal{{G}}_2(\lambda)
\end{align}
Finally, we drop the time-averaging operation, producing an error in the higher-frequency force harmonics that has zero effect on net particle motion, resulting in \eqref{Fsiggam} in the main text.

The explicit expression for the in-phase inertial force component for oscillatory flows reads:
\begin{align}
    \mathcal{G}_1=&e^{-i \sqrt{\bar{\lambda }}} \bigg[225 e^{3 \sqrt{\bar{\lambda }}} \bar{\lambda }^{3/2} \bigg(e^{2 i \sqrt{\bar{\lambda }}} \left((3+2 i) \sqrt{\bar{\lambda }}+2 i\right) \left(\text{Ei}\left((-3-i) \sqrt{\bar{\lambda }}\right)+i \pi \right)\nonumber\\
    -&\left(2+(2+3 i) \sqrt{\bar{\lambda }}\right) \left(\pi +i \text{Ei}\left((-3+i) \sqrt{\bar{\lambda }}\right)\right)\bigg)\nonumber\\
    +&48 e^{(2+i) \sqrt{\bar{\lambda }}} \left(2 \bar{\lambda }+12 \sqrt{\bar{\lambda }}+11\right) \bar{\lambda }^{5/2} \text{Ei}\left(-2 \sqrt{\bar{\lambda }}\right)\nonumber\\
    -&e^{\sqrt{\bar{\lambda }}} \left(2 \sqrt{\bar{\lambda }}+3\right) \bar{\lambda }^2 \bigg(e^{2 i \sqrt{\bar{\lambda }}} \bigg(2 \left(\sqrt{\bar{\lambda }}+(2+i)\right) \nonumber\\
    &\sqrt{\bar{\lambda }} \left(2 \bar{\lambda }+(3+3 i) \sqrt{\bar{\lambda }}+(3+6 i)\right)+15 i\bigg)\left(\pi -i \text{Ei}\left((-1-i) \sqrt{\bar{\lambda }}\right)\right)\nonumber\\
    +&\left(2 \left(\sqrt{\bar{\lambda }}+(2-i)\right) \sqrt{\bar{\lambda }} \left(2 \bar{\lambda }+(3-3 i) \sqrt{\bar{\lambda }}+(3-6 i)\right)-15 i\right) \nonumber\\
    &\left(\pi +i \text{Ei}\left((-1+i) \sqrt{\bar{\lambda }}\right)\right)\bigg)\nonumber\\
    +&e^{i \sqrt{\bar{\lambda }}} \bigg(302 \bar{\lambda }^{3/2}+144 \bar{\lambda }^{5/2}+12 \bar{\lambda }^{7/2}+8 \bar{\lambda }^4-8 \bar{\lambda }^3+36 \bar{\lambda }^2-598 \bar{\lambda }-512 \sqrt{\bar{\lambda }}\nonumber\\
    -&189\bigg)\bigg]
    \bigg/\left(160 \left(2 \bar{\lambda }^{3/2}+2 \bar{\lambda }+\sqrt{\bar{\lambda }}\right)\right),\label{G1full}
\end{align}
where $\bar{\lambda}=3\lambda/2$ and Ei is the exponential integral function. The expression for the out-of-phase component $\mathcal{G}_2$ is similarly explicit and lengthy:
\begin{align}
    \mathcal{G}_2=&e^{-i \sqrt{\bar{\lambda }}} \sqrt{\bar{\lambda }} \bigg[-240 e^{(2+i) \sqrt{\bar{\lambda }}} \left(2 \bar{\lambda }^{3/2}+6 \bar{\lambda }+6 \sqrt{\bar{\lambda }}+3\right) \bar{\lambda }^{3/2} \text{Ei}\left(-2 \sqrt{\bar{\lambda }}\right)\nonumber\\
    +&225 e^{3 \sqrt{\bar{\lambda }}} \bar{\lambda }^{3/2} \bigg(\left(3+(3+2 i) \sqrt{\bar{\lambda }}\right) \left(\text{Ei}\left((-3+i) \sqrt{\bar{\lambda }}\right)-i \pi \right)\nonumber\\
    +&e^{2 i \sqrt{\bar{\lambda }}} \left((2+3 i) \sqrt{\bar{\lambda }}+3 i\right) \left(\pi -i \text{Ei}\left((-3-i) \sqrt{\bar{\lambda }}\right)\right)\bigg)\nonumber\\
    +&e^{\sqrt{\bar{\lambda }}} \left(2 \sqrt{\bar{\lambda }}+3\right) \bar{\lambda }^2 \bigg(\left((10+14 i) \bar{\lambda }^{3/2}+4 i \bar{\lambda }^2+(30+12 i) \bar{\lambda }+30 \sqrt{\bar{\lambda }}+15\right)\nonumber\\
    &\left(\pi +i \text{Ei}\left((-1+i) \sqrt{\bar{\lambda }}\right)\right)\nonumber\\
    +&e^{2 i \sqrt{\bar{\lambda }}} \left(15-2 i \left(\sqrt{\bar{\lambda }}+(2+i)\right) \sqrt{\bar{\lambda }} \left(2 \bar{\lambda }+(3+3 i) \sqrt{\bar{\lambda }}+(3+6 i)\right)\right)\nonumber\\
    &\left(\pi -i \text{Ei}\left((-1-i) \sqrt{\bar{\lambda }}\right)\right)\bigg)\nonumber\\
    -&e^{i \sqrt{\bar{\lambda }}} \bigg(42 \bar{\lambda }^{3/2}+340 \bar{\lambda }^{5/2}+60 \bar{\lambda }^{7/2}+8 \bar{\lambda }^4+128 \bar{\lambda }^3+666 \bar{\lambda }^2-288 \bar{\lambda }+54 \sqrt{\bar{\lambda }}\nonumber\\
    +&45\bigg)\bigg]
    \bigg/\left(240 \left(2 \bar{\lambda }^{3/2}+2 \bar{\lambda }+\sqrt{\bar{\lambda }}\right)\right)\label{calG2}
\end{align}

\section{Evaluation of the memory integral and time-scale separation}\label{appendix:timescale}
We first comment on the contribution due to the history term. It is well-known that the Basset history integral poses a special challenge (cf.\ \cite{michaelides1992novel,van2011efficient,prasath2019accurate}): Its evaluation is often computationally intensive since one has to numerical solve an integro-differential equation. However, for oscillatory flows it can be evaluated explicitly---reducing to a simple ODE---and results in sub-dominant corrections to the Stokes drag and added mass forces (cf. \cite{landau1959course,danilov2000mean}), i.e., 
\begin{align}
    6\pi^{1/2} \nu^{1/2} a_p^2 \rho_f &\int_{-\infty}^t \frac{d/d\tau \left[\bm{U}_p(t) - \bm{U}(\bm{r}_p(t),t)\right]}{\sqrt{t-\tau}}d\tau \nonumber\\
    &=  \frac{1}{2}m_f \frac{d}{dt}\left[\bm{U}_p(t) - \bm{U}(\bm{r}_p(t),t) \right] \left(3\sqrt{\frac{3}{2\lambda}}\right)\nonumber\\
    &+6\pi \rho_f \nu a_p \left[\bm{U}_p(t) - \bm{U}(\bm{r}_p(t),t) \right]\left(\sqrt{\frac{3\lambda}{2}}\right).\label{eqn:basset}
\end{align}
We note that these corrections apply only if the velocity difference between the particle and the fluid is oscillatory, i.e. $(\bm{U}_p(t) - \bm{U}(\bm{X}_p(t),t))\propto e^{it}$. Therefore, \eqref{eom_full_dim} cannot be easily used to describe the unsteady particle dynamics with rectified motion due to the difficulty in evaluating the memory term. 

This apparent difficulty can be resolved by exploiting the clear separation of time-scales inherent to most fast oscillatory flow setups. Assuming all parameters are $O(1)$ and $\epsilon\ll 1$, we introduce a ``slow time" $T=\epsilon^2 t$, in addition to the ``fast time" $t$. Using the following transformations,
\begin{subequations}
	\begin{alignat}{4}
	&\bm{r}_p(t) \mapsto \bm{r}_p(t,T),\\
	&\frac{d}{d t} \mapsto \frac{\partial }{\partial t} +\epsilon^2\frac{\partial }{\partial T},\\
	&\frac{d^2}{d t^2} \mapsto \frac{\partial^2 }{\partial t^2} +2\epsilon^2\frac{\partial^2 }{\partial t \partial T} +\epsilon^4\frac{\partial^2 }{\partial T^2},
	\end{alignat}
\end{subequations}
we seek a perturbation solution in the general form: $\bm{r}_p(t,T)=\bm{r}_{p_0}(t,T)+\epsilon \bm{r}_{p_1}(t,T)+\epsilon^2 \bm{r}_{p_2}(t,T)+\dots $. On separating slow and fast time-scales and separating orders of $\epsilon$, the memory term becomes:
\begin{align}
    &\int_{-\infty}^t \frac{d/d\tau \left[d\bm{r}_p(\tau)/d\tau - \epsilon \bm{u}(\bm{r}_p(\tau),\tau)\right]}{\sqrt{t-\tau}}d\tau \nonumber\\
    &= \int_{-\infty}^t \frac{\frac{\partial^2 }{\partial \tau^2} \left(\bm{r}_{p_0}(T)+\epsilon \bm{r}_{p_1}(\tau,T) \right) - \epsilon \partial_\tau(\bm{u}_{osc}+\epsilon \bm{r}_{p_1}\cdot \nabla \bm{u}_{osc})}{\sqrt{t-\tau}}d\tau \nonumber\\
    &=\epsilon \int_{-\infty}^t \frac{\partial^2_\tau \bm{r}_{p_1}(\tau)  - \partial_\tau(\bm{u}_{osc})}{\sqrt{t-\tau}}d\tau
\end{align}
The contribution due to the $\mathcal{O}(\epsilon^2)$ nonlinear forcing term $\partial_\tau(\bm{r}_{p_1}\cdot \nabla \bm{u}_{osc})$ is identically zero for oscillatory flows, after time-averaging. Additionally, the effect on the steady flow component is higher-order in $\epsilon$ and is, therefore, neglected. Thus, the main contributions due to the history integral appear as sub-dominant corrections to the Stokes drag and added mass terms, given by \eqref{eqn:basset}, at $\mathcal{O}(\epsilon)$.

We now proceed with the formal separation of timescales of \eqref{eqn:eomunsteady}. At $O(1)$,
\begin{align}
\lambda \left(\hat{\kappa}+1\right) \frac{\partial^2 \bm{r}_{p_0}}{\partial t^2} + \frac{\partial \bm{r}_{p_0}}{\partial t} = 0
\end{align}
This equation is trivially satisfied if $\bm{r}_{p_0}=\bm{r}_{p_0}(T)$; thus, the leading order particle position $\bm{r}_{p_0}$ depends only on the slow-time $T$. At $O(\epsilon)$, we obtain the following after explicitly evaluating the history integral:
\begin{align}
\lambda \left(\hat{\kappa}+d\right) \frac{\partial^2 \bm{r}_{p_1}}{\partial t^2} + c\frac{\partial \bm{r}_{p_1}}{\partial t} 
= \left\{ \lambda d \frac{\partial \bm{u}_{osc}}{\partial t} + c \bm{u}_{osc} \right\}_{\bm{r}_{p_0}},
\end{align}
where $c = \left(1+\sqrt{\frac{3\lambda}{2}}\right)$ and $d=\left(1+\sqrt{\frac{3}{2\lambda}}\right)$ encode the Basset force contributions to the Stokes drag and added mass forces respectively. Assuming fast oscillatory inviscid flow dynamics, $\bm{u}_{osc}=\bm{u}_0(\bm{r})e^{it}$ and ignoring transients, the solution at $O(\epsilon)$ is given by
\begin{subequations}
\begin{align}
\bm{r}_{p_1} &=  \int \left(\bm{u}_{osc}+\bm{w}_{osc}\right)dt\\
\bm{w}_{osc} &= -\frac{ i\lambda\hat{\kappa}}{c +i\lambda(\hat{\kappa}+d)}\bm{u}_{osc},
\end{align}\label{rp1w1osc1d}\noindent
\end{subequations}
where we make use of complex phasors. With the $\O(\epsilon)$ oscillatory particle dynamics explicitly known, we obtain at $O(\epsilon^2)$, after time averaging:
\begin{align}
\frac{d \bm{r}_{p_0}}{d T}=& \lambda \left\langle \bm{r}_{p_1}\cdot \frac{\partial \nabla \bm{u}_{osc}}{\partial t}\right\rangle +  \left\langle \bm{r}_{p_1}\cdot \nabla\bm{u}_{osc}\right\rangle +\frac{2\lambda}{3}\left\langle \bm{u}_{osc} \cdot \nabla \bm{u}_{osc}\right\rangle \nonumber\\
+& \frac{\lambda}{3}\left\langle \frac{\partial \bm{r}_{p_1}}{\partial t}\cdot \nabla \bm{u}_{osc}\right\rangle
+\frac{2\lambda}{3} \mathcal{G}_1 \left\langle \left(\frac{\partial \bm{r}_{p_1}}{\partial t}-\bm{u}_{osc}\right)\cdot \nabla\bm{u}_{osc}\right\rangle \nonumber\\
+& \frac{2\lambda}{3} \mathcal{G}_2 \left\langle \partial_t\left(\frac{\partial \bm{r}_{p_1}}{\partial t}- \bm{u}_{osc}\right)\cdot\nabla\bm{u}_{osc}\right\rangle + \frac{2\lambda}{3} \alpha^2 \left\langle \nabla \bm{u}:\nabla \nabla \bm{u} \right\rangle \mathcal{F}\nonumber\\
= & \left\langle \left(\int \bm{w}_{osc} dt\right)\cdot \nabla \bm{u}_{osc}\right\rangle -\frac{2\lambda}{3}\left\langle \bm{w}_{osc} \cdot \nabla \bm{u}_{osc}\right\rangle \nonumber\\
& +\frac{2\lambda}{3} \mathcal{G}_1 \left\langle \bm{w}_{osc}\cdot\nabla \bm{u}_{osc}\right\rangle + \frac{2\lambda}{3} \mathcal{G}_2 \left\langle \partial_t \bm{w}_{osc}\cdot \nabla \bm{u}_{osc}\right\rangle  + \frac{2\lambda}{3} \alpha^2 \left\langle \nabla \bm{u}:\nabla \nabla \bm{u} \right\rangle\mathcal{F}.
\end{align}
Inserting \eqref{rp1w1osc1d}, and evaluating the time averages results in \eqref{eom1dslowtime} in the main text.

\section{Direct Numerical Simulation Details}\label{appendix:DNS}
Here, we present the governing equations and the numerical solution strategy employed in this work. Briefly, we consider incompressible viscous fluid in an unbounded domain, $\Sigma$, with an imposed monopolar flow field. The particle is modeled as an immersed solid, which moves under the influence of the oscillatory flow field. The particle is defined with support $\Omega$ and boundary $\partial \Omega$, respectively.
Under the aforementioned conditions, the flow in the domain can be described using the incompressible Navier–Stokes equations: 
\begin{equation}
\bv{\nabla} \cdot \gv{u} = 0 ; ~~ \frac{\partial \gv{u}}{\partial t}+(\gv{u} \cdot \bv{\nabla}) \gv{u} = -\frac{\bv{\nabla} P}{\rho} + \nu \nabla^{2} \gv{u}, ~~ \gv{x} \in \Sigma \backslash \Omega \label{eqn:ns}
\end{equation}
where $\rho$, $P$, $\gv{u}$ and $\nu$ are the fluid density, pressure, velocity and kinematic viscosity, respectively. The dynamics of the fluid–solid system is coupled via the no-slip boundary condition $\gv{u} = \gv{u_s}$ on $\partial \Omega$, where $\gv{u_s}$ is the solid body velocity. The system of equations is solved using a velocity–vorticity formulation with a combination of remeshed vortex methods and Brinkmann penalization implemented in an axisymmetric solver \cite{pyaxisymflow2023}. The monopole and particle are placed on the axis of symmetry, separated by a center-to-center distance $r_{p}(0)$. The hydrodynamic forcing contributions arising from the density mismatch between the fluid and solid are accounted for via the unsteady term proposed in \cite{engels2015numerical}. The chosen computational methodology has been validated across a range of flow–structure interaction problems, from flow past bluff bodies to biological swimming, as well as for 2D and 3D streaming flows (see Refs.~ \cite{gazzola2011simulations,parthasarathy2019streaming,bhosale_parthasarathy_gazzola_2020,bhosale2022multicurvature,chan2022three,bhosale2022soft,pyaxisymflow2023} for details). The DNS code and example cases can be accessed online, see \cite{pyaxisymflow2023}.

\section{Fitting procedure to obtain $\mathcal{G}$ from DNS}\label{appendix:fit}
The DNS produces (unsteady) particle trajectories as a function of time. As depicted in Fig.~\ref{fig:DNSresults}(b) of the main text, these oscillatory trajectories were time-averaged over one period to obtain the steady particle dynamics $r_{p}(T)$, which is a function of the \textit{slow} time $T=\epsilon^2 t$. We fit these trajectories to \eqref{eom1dslowtimemonopole} in the main text with $\cal{G}$ as the fitting parameter in order to obtain the simulation points of Fig.\ref{fig:traj_Gcal}(e) of the main text. The fitting process involves the following steps: i) We first validate the DNS technique for density-matched particles using the function $\mathcal{F}$ established in \cite{agarwal2021unrecognized} for all considered $\lambda$ values, obtaining an accuracy within $5\%$ using the current DNS methodology. ii) Next, slow-time particle trajectories are obtained by numerically integrating \eqref{eom1dslowtimemonopole} in the main text, with the full analytical expression $\mathcal{F}$ from \cite{agarwal2021unrecognized}. These are fitted to the time-averaged trajectories obtained from DNS using the method of least squares, resulting in the direct determination of $\mathcal{G}$. The error bars of the fit are computed for each $\mathcal{G}$ value, assuming an error of $5\%$ in the values of $\mathcal{F}$, consistent with the maximum error observed in the density-matched validation case.

\bibliographystyle{jfm}
\bibliography{biblio}

\end{document}